\documentclass[preprintnumbers,amsmath,amssymb,floatfix,11pt,prd,onecolumn,superscriptaddress,nofootinbib]{revtex4}
\usepackage{graphicx}
\usepackage{dcolumn}
\usepackage{amsmath,amssymb,mathtools}
\usepackage{epstopdf}
\usepackage{verbatim}
\usepackage{amsmath}
\usepackage[colorlinks=true, citecolor=blue, urlcolor = blue, linkcolor= blue,
bookmarks=true]{hyperref}
\labelformat{section}{Section #1}
\labelformat{subsection}{Section #1}
\labelformat{subsubsection}{Section #1}
\labelformat{subsubsubsection}{Section #1}
\labelformat{equation}{Eq.~(#1)}
\labelformat{figure}{Fig.~#1}
\labelformat{subfigure}{Fig.~\thefigure#1}
\labelformat{table}{Tab.~#1}
\labelformat{appendix}{Appendix #1}
\begin{document}	
\title[]{The extended phase space thermodynamics and Ehrenfest scheme for the Kerr-Sen AdS black holes}

\author{Md Sabir Ali} 
\email{alimd.sabir3@gmail.com,sabirali@mahishadalrajcollege.ac.in} 
\affiliation{Department of Physics, Mahishadal Raj College, Purba Medinipur, West Bengal 721628, India}
\affiliation{Institute of Theoretical Physics \&
Research Center of Gravitation, Lanzhou University, Lanzhou 730000, China}
\affiliation{School of Physical Science and Technology, Lanzhou University, Lanzhou 730000, China}
\author{Arindam Mondal} 
\email{mondal.arindam622@gmail.com} 
\affiliation{Department of Physics, Mahishadal Raj College, Purba Medinipur, West Bengal 721628, India}
\author{Sushant G. Ghosh} \email{sghosh2@jmi.ac.in}
\affiliation{Centre for Theoretical Physics, Jamia Millia Islamia, New Delhi 110025, India}
\affiliation{Astrophysics and Cosmology Research Unit, School of Mathematics, Statistics and Computer Science, University of KwaZulu-Natal, Private Bag X54001, Durban 4000, South Africa}

%%%%%%%%%%%%%%%%%%%%%%%%%%%%%%%%%%%%%%%%%%%%%%%%%%%%%%%%%%%%%%%%%%%%%%%%%%
%%%%%%%%%%%%%%%%%%%%%%%%%%%%%%%%%%%%%%%%%%%%%%%%%%%%%%%%%%%%%%%%%%%%%%%%%%%%%%%%%%%%%%%%%%%%%%%%%%%%%%%%%%%%%%%%%%%%%%%%%%%%%%%%%%%%%%%%%%%%%%%%%%%%
\begin{abstract}
In the present work, we numerically investigate the horizon structure of the Kerr-Sen black holes in anti-de Sitter (AdS) spacetime. Further, we investigate the phase transitions and critical phenomena in Kerr-Sen-AdS black holes at the critical points. Such black holes are characterized by its mass ($M$), the dilaton charge ($Q$), and the negative cosmological constant, $\Lambda(<0)$. We define a dimensionless parameter $\epsilon=\bar{J}/{\bar{Q}^2}$ and express the mass, temperature, volume, and Gibbs free energy in terms of $\epsilon$ and its polynomials. Moreover, we numerically fit the data for the critical points and find that in the appropriate limit, the expressions for critical points would correspond to the respective critical points of the Kerr-AdS black hole thermodynamics. Such a study involves a systematic analysis of temperature, Gibbs free energy, and volume in the extended phase space. We provide an analytical verification of the nature of the phase transitions at the critical points by introducing the Ehrenfest equations. We find that the Ehrenfest equations are satisfied in accordance with the phase behaviour. We also check that all three quantities, e.g., the specific heat at constant pressure, $C_P$, the volume expansion coefficient, $\alpha$, and the isothermal compressibility, $\kappa_T$, diverge at the critical points. We find the $Prigogine$-$Defay$ ratio using the expressions of $C_P$, $\alpha$, and $\kappa_T$, and find that it identically equals unity. Hence, the phase transition behavior of the Kerr-Sen-AdS black holes at their critical points is of second order. In addition, we propose investigating the energy extraction process via the Penrose process by deriving work efficiency. Later, we calculate the speed of sound and adiabatic compressibility for the rotating Kerr-Sen-AdS black holes. Finally, on a specific note, we calculate the thermodynamic quantities of the boundary conformal field theory (CFT) dual to the extended phase space.  
\end{abstract}

\maketitle
\tableofcontents
\section{Introduction}
The thermodynamics of black holes has profound effects on the assimilation of ideas connecting gravitation, thermodynamics, and quantum field theory. To understand the concepts of quantum gravity, black hole thermodynamics may play a crucial role in this understanding. The laws of black hole mechanics are in complete analogy with the laws of an ordinary thermal system. The four laws of black hole mechanics were discovered in the pioneering work by Bardeen, Carter, and Hawking \cite{Bardeen1973}. Bekenstein, in his subsequent analysis, showed that the entropy has a very close connection to the area of the black hole event horizon \cite{Bekenstein:1973ur}. At this juncture, it is to be mentioned that even though the black holes follow the thermal properties, they are not stable if we consider the case of Schwarzschild black holes. To be more specific, the thermodynamics of the Schwarzschild black holes in AdS spacetime has its deep-rooted meaning in understanding the notion of quantum gravity because of its relevance to AdS/CFT correspondence \cite{Maldacena:1997re}. This profound idea was first realised when Hawking and Page observed that a phase transition exists between the Schwarzschild-AdS and thermal AdS spaces, commonly known as the Hawking-Page phase transition \cite{Hawking1983}. The Schwarzschild-AdS black hole is thermally stable. Later, such findings have triggered research on black hole thermodynamics in AdS spacetime for various solutions, both in general and in modified theories of gravity. These studies also consider the variation of the negative cosmological constant $(\Lambda < 0)$ as a thermodynamic variable, identified as the thermodynamic pressure ($P$) and its corresponding conjugate quantity, the thermodynamic volume ($V$). Inclusion of the $PdV$ term in the first law is consistent with the first law of black hole mechanics, and the corresponding Smarr-Gibbs-Duhem relation or the extended Euler relation equally holds \cite{Kubiznak:2016qmn,Mann:2015luq,Dolan:2012jh, Dolan:2011xt}. This gives rise to the extended phase space thermodynamics of black holes in asymptotically AdS spacetimes, known as extended black hole mechanics/black hole chemistry, where the concept of enthalpy is elaborated \cite{Kastor:2009wy, Dolan:2010ha, Cvetic:2010jb, Dolan:2014jva}. A careful analysis of the charged AdS black holes shows that the small-large phase transitions at constant electric charge have a strong resemblance to the famous van der Waals (vdW)-like phase transition of the ordinary thermodynamic systems \cite{Kubiznak:2012wp,Dehyadegari:2017flm,Hennigar:2016xwd}. A similar analysis for the Kerr-AdS or the Kerr-Newman-AdS shows a strong analogy with the van der Waals (vdW) fluids \cite{Banerjee:2010bx, Altamirano:2013uqa, Altamirano:2014tva, Caldarelli:1999xj}. The same conditions as those for other AdS black holes are also met. Among them are the AdS black holes in higher dimensions and several other modified theories of gravity \cite{Cai:2013qga,Banerjee2011,Xu:2013zea,Mo:2014qsa,Mo:2014mba,Wei:2015ana,Xu:2014tja,Zou:2014mha}. In such cases, certain differences in the phase structure and a potential deviation from the universal ratio as well as the $P$-$V$ criticality are observed. Other studies of the phase structure in the extended phase space are analysed in the nonlinear electrodynamics, in $f(R)$ gravity theory, and many other gravity theories \cite{Ali:2019myr, Hendi:2012um, Chen:2013ce, Zhao:2013oza,Sherkatghanad:2014hda,Xu:2015rfa, Mirza:2014xxa}. Such studies invoke some exotic and interesting phenomena apart from the usual vdWs liquid-vapor phase transition.\\
Along with vdW fluid-like behaviour, many black hole systems in AdS spacetimes exhibit even richer phase structures, as observed in gels and polymers, commonly termed glassy phase transitions, where the Prigogine-Defay ratio deviates from unity \cite{Moynihan:1976Mo, Moynihan:1981Mo, Banerjee:2010qk}. Some such systems exhibit multiple critical phenomena, such as reentrant phase transitions, isolated critical points, and triple point structures \cite{Dolan:2014vba,Wei:2014hba,Frassino:2014pha,Altamirano:2013ane,Ali:2023wkq,Zou:2013owa,Gunasekaran:2012dq,Yang:2021ljn}. These behaviors are observed in a limited number of AdS black holes systems. In most cases, the thermodynamics of AdS black hole systems mimic the vdW fluid. The AdS black holes show the oscillatory pattern in the isotherms and the shallow tail behaviour of the Gibbs free energy, as well as the mean field theoretic constructions such as the critical exponents and consequently the scaling properties around the critical points, as we observe among the prominent properties of the vdW fluids. Although the analytical descriptions of such behaviours are limited only to a few AdS black hole systems, we can still examine them through numerical techniques. \\

Another interesting phenomenon that emerges in AdS spacetimes during the black hole phase transition is the existence of a coexistence line between the small and large black hole phases. Coexistence lines of the black hole phases are determined from the oscillatory behaviour of the isothermal curve or isobaric curve \cite{Lan:2015bia,Spallucci:2013osa} or from the intersection points of the Gibbs free energy versus the temperature diagrams. The critical points are determined through the point of inflexion in the isothermal/isobaric curves by applying the Maxwell equal-area law. The coexistence lines are then determined at the critical points. Numerous studies are devoted to this direction \cite{Spallucci:2013osa,Lan:2015bia,Spallucci:2013jja}. \\

Our paper primarily focuses on the issues of the horizon structure and thermodynamic quantities in the extended phase space of the Kerr-Sen AdS black holes. We numerically carried out the analysis of the critical points and phase structure for Kerr-Sen-AdS black holes. We present the limiting values of the critical parameters corresponding to Kerr-AdS black holes. Later, using critical point analysis, we constructed phase diagrams and Ehrenfest schemes at the critical points. Such studies are important in their own right.  
\section{A Brief overview of the Kerr-Sen-AdS black hole solution}
%%%%%%%%%%%%%%%%%%%%%%%%%%%%%%%%%%%%%%%%%%%%%%%%%
In this section, we provide a brief review of the Lagrangian density for the gauged Einstein-Maxwell-Dilaton-Axion theory in asymptotically AdS spacetimes, incorporating a nonzero negative cosmological constant. The Lagrangian density of such a theory is written as \cite{Ali:2023ppg, Wu:2020cgf, Wu:2020mby} 
\begin{eqnarray}
\bar{\mathcal{L}}=\sqrt{-g}\left[R-\frac{1}{2}(\partial\bar{\phi})^2
 -\frac{1}{2}e^{2\bar{\phi}}(\bar{\chi})^2-e^{-\bar{\phi}}F^2 +\frac{1}{l^2}\big[4+e^{-\bar{\phi}}+e^{\bar{\phi}}(1 +\bar{\chi}^2)\big]\right]
 +\frac{\bar{\chi}}{2}\epsilon^{\mu\nu\rho\lambda}F_{\mu\nu}F_{\rho\lambda},
\end{eqnarray}
where $l$ is  the curvature radius and related to the negative cosmological constant via $l^2=-\frac{3}{\Lambda}$, $R$ is the Ricciscalar, $\bar{\phi}$ is the dilaton scalar field for the rotating case, $F_{\mu\nu}$ is the electromagnetic field tensor, and its invariant $F^2=F_{\mu\nu}F^{\mu\nu}$, $\bar{\chi}$ is the pseudo scalar axion field dual to the three-form antisymmetric tensor: $H=e^{-2\bar{\phi}}\star d\bar{\chi}$, $H^2=H_{\mu\nu\rho}H^{\mu\nu\rho}$ and $\varepsilon^{\mu\nu\rho\lambda}$ is the antisymmetric tensor called the Levi-Civita tensor density. As in the case of the ungauged case, the above Lagrangian comprises a potential term involving the dilation and axion fields.\\

\begin{figure}
\centering
    \includegraphics[scale=0.85]{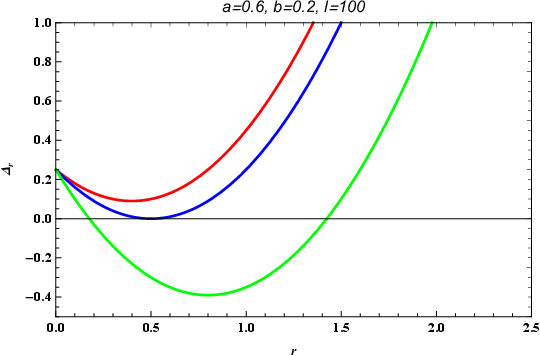}\hspace{0.35cm}
    \includegraphics[scale=0.85]{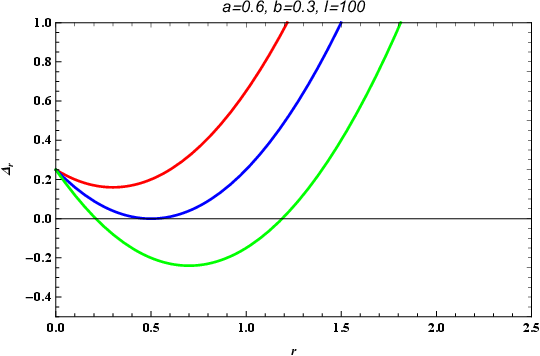}\\
    \caption{Plot showing the behavior of $\Delta_r$ vs the radial coordinate $r$ for different values $a,\,b\,\text{and}\,l$. We have two distinct horizons for $m>m_{\text{ext}}$. The red solid curve represents the naked singularity when the mass parameter $m<m_{\text{ext}}$ while the blue curve represents the extremal case, $m=m_{\text{ext}}$ when the two horizons coincide. }
    \label{fig:horizon}
\end{figure}

The four-dimensional Kerr-Sen AdS black hole was obtained in \cite{Ali:2023ppg, Wu:2020cgf, Wu:2020mby}, which, in the usual Boyer-Lindquist coordinates, $(t,r,\theta,\phi)$, is written as
exquisite forms:
\begin{eqnarray}
\label{KSAdS}
d\bar{s}^2 &=& -\frac{\Delta_r}{\Sigma}\left(dt -\frac{a\sin^2\theta}{\Xi}d\phi\right)^2
 +\frac{\Sigma}{\Delta_r}dr^2 +\frac{\Sigma}{\Delta_\theta}d\theta^2 +\frac{\Delta_\theta\sin^2\theta}{\Sigma}\left(adt
 -\frac{r^2+2br+a^2}{\Xi}d\phi\right)^2\quad\quad \,  \\
\bar{A} &=& \frac{qr}{\Sigma}\left(dt -\frac{a\sin^2\theta}{\Xi}d\phi\right) \,  \nonumber\\
\bar{\phi} &=& \ln\Big(\frac{r^2 +a^2\cos^2\theta}{\Sigma}\Big) \, , \quad
\bar{\chi} = \frac{2ba\cos\theta}{r^2 +a^2\cos^2\theta} \,\,\, ,
\end{eqnarray}

where $\Sigma = r^2 +2br +a^2\cos^2\theta$ , and the quantities
$$
\Delta_r=\Big(1 +\frac{r^2 +2br}{l^2}\Big)(r^2 +2br +a^2) -2mr, \, $$
and $$\Delta_\theta= 1 -\frac{a^2}{l^2}\cos^2\theta,\quad
\Xi = 1 -\frac{a^2}{l^2}. \,$$
This metric has a similar structure to the Kerr/Kerr-Newman AdS spacetimes. Since it is an axisymmetric metric having two Killing vectors $\xi^{\mu}_{(t)}$ and $\chi^{\mu}_{(\phi)}$. The vector $\xi^{\mu}_{(t)}$ corresponds to the time translation symmetry, while the vector $\chi^{\mu}_{(\phi)}$ corresponds to the axial symmetry. The Killing vector $\xi^{\mu}_{(t)}$ is chosen in such a way 
that at the asymptotic infinity (i.e., $r\to\infty$), it represents the timelike trajectories for the Lorentz coordinate system as confirmed from the asymptotic form of the metric \ref{KSAdS}. Therefore, we separate out the timelike Killing field $\xi^{\mu}_{(t)}$ that corresponds to the case when there is no rotation about the symmetry axis at the asymptotically large distances \cite{Frolov:1998wf}. Consequently, we could also single out the rotational Killing field $\chi^{\mu}_{(\phi)}$ corresponding to the condition that its integral lines are closed. Since the metric is not symmetric only under $t\to -t$, reflecting the fact that the Kerr-Sen-AdS black holes are not static but stationary. The linear combination, $$\zeta^{\mu}=\xi^{\mu}_{(t)}+\Omega_h\chi^{\mu}_{(\phi)},$$ of the Killing vector fields $\xi^{\mu}_{(t)}$ and $\chi^{\mu}_{(\phi)}$ is also a Killing field, where $\Omega_h$ is the angular velocity at the event horizon and of course, $\zeta^{\mu}$ satisfies the Killing equation. The various metric coefficients $g_{tt},\, g_{t\phi}\, \text{and}\, g_{\phi\phi} $ are determined from the scalar products of $\xi^{\mu}_{(t)}$ and $\chi^{\mu}_{(\phi)}$, such that
$$g_{tt}=\xi_{(t)}\cdot \xi_{(t)}=\frac{a^2\sin^2\theta\Delta_\theta-\Delta_r}{\Sigma},$$$$g_{t\phi}=\xi_{(t)}\cdot \xi_{(\phi)}=\frac{a(\Delta_r-(r^2+2br+a^2)\Delta_\theta)\sin^2\theta}{\Sigma\Xi},$$
$$g_{\phi\phi}=\xi_{(\phi)}\cdot \xi_{(\phi)}=\frac{((r^2+2br+a^2)^2\Delta_\theta-a^2\Delta_r\sin^2\theta)\sin^2\theta}{\Sigma\Xi^2}.$$
\\
Next, we discuss the horizon structure of the rotating Kerr-Sen-AdS black holes \ref{KSAdS}. The Kerr-Sen-AdS black holes are characterized by the parameters $m$ (mass) $a$ (spin), $b$ (diatonic charge), and the additional parameter $l$ (the radius of the curvature).  At the event horizon, we get a well-defined null hypersurface, meaning that the normal to this surface is null. A careful calculation shows that at the event horizon, $g^{rr}=0$ $\Rightarrow \Delta_r=0$, thereby determining the size of the event horizon. The equation $\Delta_r=0$, has two possible real roots and has the analytical form. Let $r=r_h$ be the black hole event horizon and $r=r_c$ be the inner horizon or the Cauchy horizon. At $r_h=r_c$, two horizons coincide.  This is called the extremality condition, and in addition to $\Delta_r=0$, its derivative $\Delta_r^{\prime}=0$, there as well. For the complexity of the form of the horizons, we instead of writing their analytical form, we plot them in \ref{fig:horizon}. It is shown that there exist two distinct horizons for $m>m_{\text{ext}}$, and no horizons when $m<m_\text{ext}$. Two distinct horizons coincide at $m=m_{ext}$, corresponding to the extremal black hole. The quantity $m_\text{ext}$ denotes the mass of the extremal Kerr-Sen-AdS black holes. \\ 

Now, we concentrate on the discussion of the angular velocity of the Kerr-Sen-AdS black holes. To have a physical viewpoint, we consider a stationary observer moving around $\phi$-direction with the angular velocity, $\Omega=\frac{d\phi}{dt}$, which can take any arbitrary but uniform value. The four-velocity of the observer is given by $u^\mu=\gamma \zeta^\mu$, where $\gamma$ is a normalisation constant. The norm of the four velocity is given as $u^\mu u_\mu=\gamma^2(g_{tt}+2\Omega g_{t\phi}+\Omega^2 g_{\phi\phi})$. For $u^\mu$ to be timelike we have $u^\mu u_\mu=-1$, and hence $\gamma^{-2}=-g_{\phi\phi}\left({g_{tt}}/{g_{\phi\phi}}+2{g_{t\phi}}/{g_{\phi\phi}}\Omega+\Omega^2\right)$. Since $\zeta^\mu$ has to be timelike, it is impossible to have stationary observer anywhere in the Kerr-Sen-AdS spacetimes. Therefore, the requirement $\gamma^{-2}>0$ leads us to have the value of the angular velocity $\Omega_-<\Omega<\Omega_+$, where $$\Omega_\pm=\omega\pm \frac{\Xi\sqrt{\Delta_r}\Delta_\theta \Sigma \csc\theta}{(r^2+2br+a^2)^2\Delta_\theta-\Delta_r\sin^2\theta},\,\,\, \omega=\frac{a((r^2+2br+a^2)\Delta_\theta-\Delta_r)\Xi}{(r^2+2br+a^2)^2\Delta_\theta-\Delta_r\sin^2\theta}.$$
At the horizon $\Delta_r=0$, and the angular velocity reduces to the form $\Omega_\pm=\omega$ \cite{Wu:2020cgf}, where $$\Omega_-=\Omega_+\equiv\Omega_h=\omega(r_h)=\frac{a\Xi}{r_h^2+2br_h+a^2}$$
The angular velocity $\Omega_h$ at the asymptotic infinity is derived in the static frame. But at the asymptotic infinity we always have a non-vanishing angular velocity, $\omega(\infty)=\Omega_\infty=-\frac{a}{l^2}$ for the rotating AdS black holes, and hence we cannot have a static frame at all but rather a rotating one. The thermodynamically important angular velocity that is taken into account in the first law of black hole mechanics is written as
$$\Omega=\Omega_h-\Omega_\infty=\frac{a\left(1+\frac{r_h^2+2br_h}{l^2}\right)}{r_h^2+2br_h+a^2}.$$
\begin{figure}
    \centering
    \includegraphics[scale=0.8]{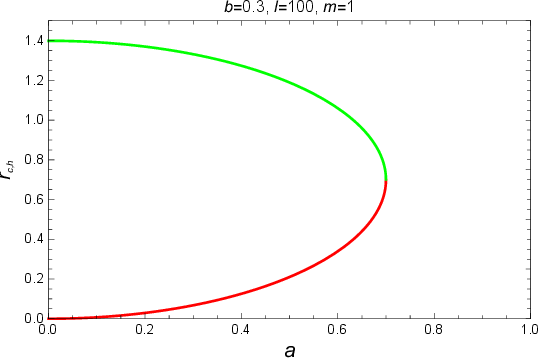}\hspace{0.45cm}
    \includegraphics[scale=0.8]{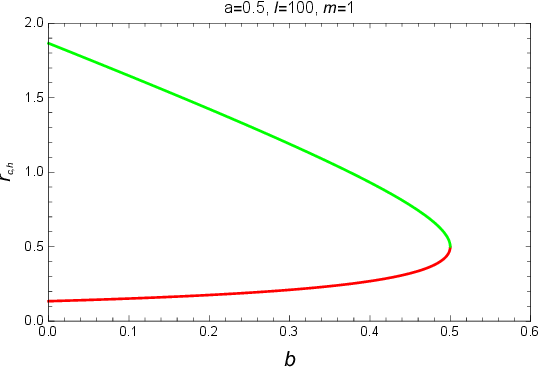}\\
    \includegraphics[scale=0.8]{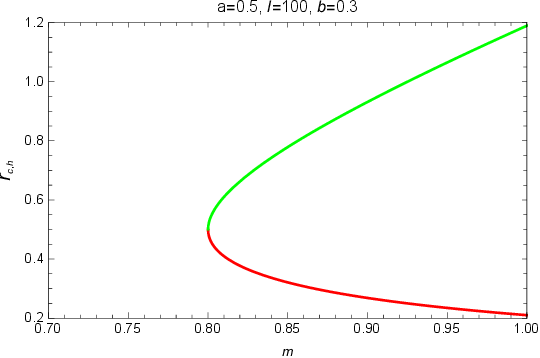}\hspace{0.45cm}
    \includegraphics[scale=0.8]{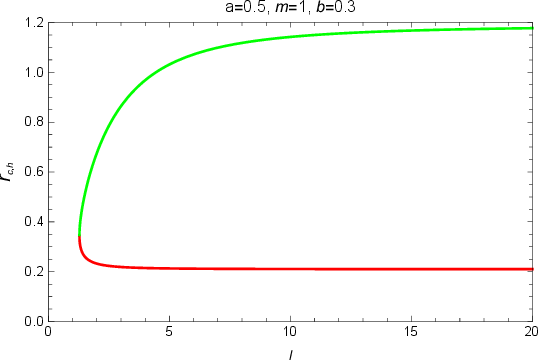}\\
    \caption{Plot showing the behavior of the Cauchy horizon ($r_c$) and event horizon ($r_h$) vs the rotation parameter $a$ and the parameter $b$ (upper panel) and the mass parameter $m$ and the AdS radius $l$ (lower panel). We plot them for various parametric values of $a,\,b,\,l\,\text{and}\,\,m$.}
    \label{fig:horizon_1}
\end{figure}
We depict in the \ref{fig:horizon_1}, the black hole event horizon $r_h$ and the Cauchy horizon $r_c$ for a parametric set of values of the various parameters.
\section{Thermodynamics of Kerr-Sen-AdS black hole}
This section is devoted to the discussion of the thermodynamic phase behaviour of the Kerr-Sen-AdS black hole. For the Killing vector $\zeta^\mu$, the surface gravity $\kappa$ is given as
\begin{eqnarray}
\label{suface_gravity}
 \kappa^2=-\frac{1}{2}\zeta^{\mu;\nu}\zeta_{\mu;\nu}.    
\end{eqnarray}
The temperature $\bar{T}$ of the black hole event horizon is related to the surface gravity via $\bar{T}=\frac{\kappa}{2\pi}$. Further, the mass parameter of the black hole, its value at the event horizon is calculated from $\Delta_r=0$. Having defined the temperature and the mass parameter, we are now in a position to delve into the details of the thermodynamic quantities of the Kerr-Sen-AdS black hole. In the presence of the negative cosmological constant, we derive the thermodynamic quantities associated with the solution \ref{KSAdS} which are computed through the standard method and therefore they have the following expressions \cite{Ali:2023ppg, Wu:2020cgf, Wu:2020mby}:
\begin{eqnarray}\label{Therm}
&&\bar{M}= \frac{m}{\Xi^2} \, , \quad \bar{J} = \frac{ma}{\Xi^2} \, , \quad
\bar{Q} = \frac{q}{\Xi} \, , \\
&&\bar{T} = \frac{(r_+ +b)(2r_+^2 +4br_+ +l^2 +a^2)
 -ml^2}{2\pi(r_+^2 +2br_+ +a^2)l^2} \, , \\
&&\bar{S} = \frac{\pi(r_+^2 +2br_+ +a^2)}{\Xi} \, , \quad
 \bar{\Omega} = \frac{a\Xi}{r_+^2 +2br_+ +a^2} \, , \\
\label{entropy}
&&\bar{\Phi} = \frac{qr_+}{r_+^2 +2br_+ +a^2} \, .
\end{eqnarray}

The above thermodynamic quantities follow the extended Smarr relation and are analysed through the dimensional analysis using Euler's formula \cite{Wu:2020cgf, Wu:2020mby} 
\begin{eqnarray}
\bar{M} = 2\bar{T}\bar{S} +2\bar{\Omega}\bar{J} +\bar{\Phi}\bar{Q} -2\bar{V}\bar{P} \, .
\end{eqnarray}
The thermodynamic volume $\bar{V}$ in such a case is derived to be 
\begin{eqnarray}
\label{volume_rh}
\bar{V} = \frac{4\pi}{3\Xi}(r_+ +b)(r_+^2 +2br_+ +a^2) \, ,
\end{eqnarray}
and conjugate to the volume, the pressure term is calculated to be $\bar{P} = 3/(8\pi\,l^2)$. Unfortunately, the first law of the Kerr-Sen-AdS black hole mechanics is written to be \cite{Wu:2020cgf, Wu:2020mby} 
\begin{eqnarray}
d\bar{M} = \bar{T}d\bar{S} +\bar{\Omega}d\bar{J} +\bar{\Phi}d\bar{Q}
 +\bar{V}d\bar{P} +\bar{J}d\Xi/(2a) \, .
\end{eqnarray}
The reason behind such ambiguity comes from the fact that we adopted the frame at infinity, which is rotating. Such ambiguity could be avoided if we would define a frame at infinity which is at rest. In the subsequent discussion, we address such issues of defining the rest frame at asymptotic infinity. 

\begin{figure}
    \includegraphics[width=7.5cm,height=6.cm]{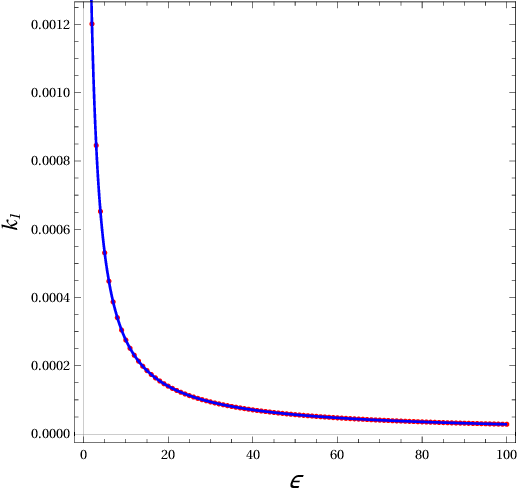}\hspace{0.45cm}
    \includegraphics[width=7.5cm,height=6.cm]{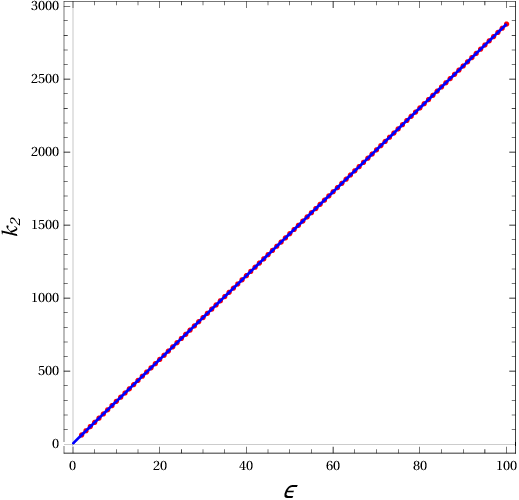}\\
    \includegraphics[width=7.5cm,height=6.cm]{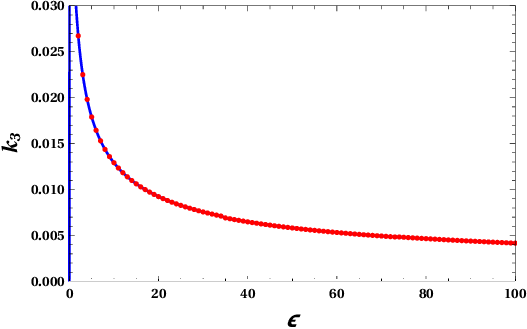}\hspace{0.45cm}
    \includegraphics[width=7.5cm,height=6.cm]{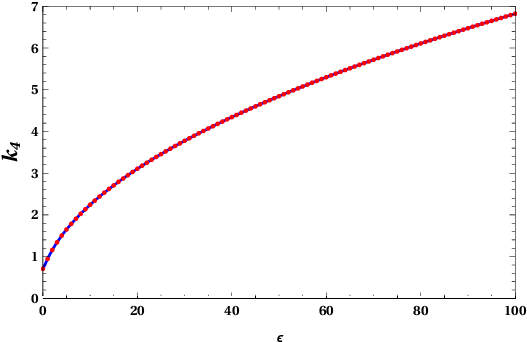}\\
    \includegraphics[width=7.5cm,height=6.cm]{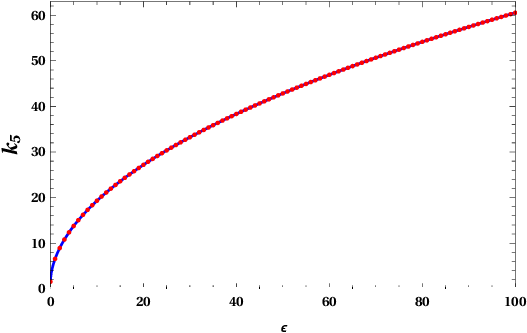}\hspace{0.45cm}
    \caption{Plot showing the behaviour of the coefficients $k_1$, $k_2$, $k_3$, $k_4$, and $k_5$ as a function of $\epsilon$. The dots represent the exact numerical values, and the solid lines denote their corresponding fitting results.}
    \label{Error}
\end{figure}

We take a simple coordinate transformation $\phi\rightarrow \phi -a\,t/l^2$, which could transform the Kerr-Sen-AdS black holes into a frame that is rest at the asymptotic infinity. A careful calculation of the thermodynamic quantities in such a rest frame could help us to define the thermodynamic important quantities where the changes in the mass, the angular velocity and the thermodynamic volume are observed. These thermodynamic quantities  given are written in \ref{Therm} in their modified forms as \cite{Wu:2020cgf, Wu:2020mby} 
\begin{eqnarray}
\widetilde{M} = \bar{M} +\frac{a}{l^2}\bar{J} \, , \quad
 \widetilde{\Omega} = \bar{\Omega} +\frac{a}{l^2} \, ,
\quad \widetilde{V} = \bar{V} +\frac{4\pi}{3}a\bar{J} \, .
\end{eqnarray}
Now, we are in a situation to find the newly defined thermodynamic quantities which indeed satisfy the standard
form of the first law and consequently the Smarr mass formula \cite{Wu:2020cgf, Wu:2020mby},
\begin{eqnarray}
d\widetilde{M} &=& \bar{T}d\bar{S} +\widetilde{\Omega} d\bar{J}
 +\bar{\Phi}d\bar{Q} +\widetilde{V}d\bar{P} \, , \\
\widetilde{M} &=& 2\bar{T}\bar{S} +2\widetilde{\Omega}\bar{J}
 +\bar{\Phi}\bar{Q} -2\widetilde{V}\bar{P} \, .
\end{eqnarray}
The newly defined mass parameter of the Kerr-Sen-AdS black holes in terms of $\bar{J}$, $\bar{S}$ and $\bar{P}$ is given as
\begin{eqnarray}
\label{massfunction}
\widetilde{M}=\frac{12 \pi ^2 \bar{J}^2 (8 \bar{P} \bar{S}+3)+96 \pi ^2 \bar{J} \bar{P} \bar{S}+\bar{S} (8 \bar{P} \bar{S}+3) \left(\bar{S} (8 \bar{P} \bar{S}+3)+6 \pi  \bar{Q}^2\right)}{6 \sqrt{\pi } \sqrt{S} \sqrt{(8 \bar{P} \bar{S}+3) \left(12 \pi ^2 \bar{J}^2+\bar{S} \left(\bar{S} (8 \bar{P} \bar{S}+3)+6 \pi  \bar{Q}^2\right)\right)}},
\end{eqnarray}
In terms of the angular momentum $\bar{J}$, the entropy $\bar{S}$ and the pressure $\bar{P}$, the temperature $\bar{T}$, Gibbs free energy $\bar{G}$, and the volume $\bar{V}$ are expressed to be
\begin{eqnarray}
\label{tempads}
\bar{T} &=& \frac{\bar{S}^2 \left(16 \bar{P} \left(\pi  \bar{Q}^2+2 \bar{S}\right)+64 \bar{P}^2 \bar{S}^2+3\right)-12 \pi ^2 \bar{J}^2}{4 \sqrt{\pi } \bar{S}^{3/2} \sqrt{\left(8 \bar{P} \bar{S}+3\right) \left(12 \pi ^2 \bar{J}^2+\bar{S} \left(\bar{S}\left(8 \bar{P} \bar{S}+3\right)+6 \pi  \bar{Q}^2\right)\right)}},\\
\label{Gibbs}
\bar{G} &=& \frac{12 \pi ^2 \bar{J}^2 \left(16 \bar{P} \bar{S}+9\right)+\bar{S} \left(12 \pi  \bar{Q}^2 \left(4 \bar{P} \bar{S}+3\right)-64 \bar{P}^2 \bar{S}^3+9 \bar{S}\right)}{12 \sqrt{\pi } \sqrt{\bar{S}} \sqrt{\left(8 \bar{P} \bar{S}+3\right)\left(12 \pi ^2 \bar{J}^2+\bar{S} \left(\bar{S} \left(8 \bar{P} \bar{S}+3\right)+6 \pi  \bar{Q}^2\right)\right)}},\,\\
\bar{V}&=&\frac{4 \sqrt{S} \left(6 \pi ^2 \bar{J}^2+\bar{S} \left(\bar{S} (8 \bar{P} \bar{S}+3)+3 \pi  \bar{Q}^2\right)\right)}{3 \sqrt{\pi } \sqrt{(8 \bar{P} \bar{S}+3) \left(12 \pi ^2 \bar{J}^2+S \left(\bar{S} (8 \bar{P} \bar{S}+3)+6 \pi  \bar{Q}^2\right)\right)}}.
\label{Volume}
\end{eqnarray}
The critical points are determined through the conditions $\partial_{\bar{S}}\bar{T}=\partial_{\bar{S},\bar{S}}\bar{T}=0$ \cite{Cheng:2016bpx}, which leads us to have
\begin{eqnarray}
\label{pt1}
&&144 \pi ^4 \bar{J}^4 (32 \bar{P} \bar{S}+9)+24 \pi ^2 \bar{J}^2 \bar{S} \left(36 \pi  \bar{Q}^2 (4 \bar{P}\bar{S}+1)+\bar{S}(8 \bar{P}\bar{S}+3)^2 (16 \bar{P} \bar{S}+3)\right)+\bar{S}^4 \big(4096 \bar{P}^4 \bar{S}^4+\nonumber\\
&&2048 \bar{P}^3 \bar{S}^2 \left(3 \pi  \bar{Q}^2+2 \bar{S}\right)-384 \bar{P}^2 \left(2 \pi ^2 \bar{Q}^4-12 \pi  \bar{Q}^2 \bar{S}-3 \bar{S}^2\right)+864 \pi  \bar{P} \bar{Q}^2-27\big)=0,
\end{eqnarray}
\begin{eqnarray}
&&-5184 \pi ^6 \bar{J}^6 \left(512 \bar{P}^2 \bar{S}^2+288 \bar{P} \bar{S}+45\right)-144 \pi ^4 \bar{J}^4 \bar{S} \big(72 \pi  \bar{Q}^2 \left(320 \bar{P}^2 \bar{S}^2+174 \bar{P} \bar{S}+27\right)+\bar{S}\nonumber\\
&&\big(-32768 \bar{P}^4 \bar{S}^4+20160 \bar{P}^2 \bar{S}^2+8640 \bar{P} \bar{S}+1053\big)\big)-12 \pi ^2 \bar{J}^2 \bar{S}^2 \big(2592 \pi ^2 \bar{Q}^4 \big(40 \bar{P}^2 \bar{S}^2+20 \bar{P} \bar{S}+3\big)\nonumber\\
&&-24 \pi  \bar{Q}^2 \bar{S} \big(4096 \bar{P}^4 \bar{S}^4-9984 \bar{P}^3 \bar{S}^3-11520 \bar{P}^2 \bar{S}^2-3780 \bar{P} \bar{S}-405\big)+5 \bar{S}^2 (8 \bar{P} \bar{S}+3)^4 (32 \bar{P} \bar{S}+9)\big)\nonumber\\
&&+\bar{S}^6 \big(-262144 \bar{P}^6 \bar{S}^6-196608 \bar{P}^5 \bar{S}^4 \left(5 \pi  \bar{Q}^2+2 \bar{S}\right)-184320 \bar{P}^4 \bar{S}^2 \left(-4 \pi ^2 \bar{Q}^4+8 \pi  \bar{Q}^2 \bar{S}+\bar{S}^2\right) \nonumber\\
&&+55296 \pi  \bar{P}^3 \bar{Q}^2 \big(2 \pi ^2 \bar{Q}^4+8 \pi  \bar{Q}^2 \bar{S}-15 \bar{S}^2\big)+5184 \bar{P}^2 \left(12 \pi ^2 \bar{Q}^4-40 \pi  \bar{Q}^2 \bar{S}+5 \bar{S}^2\right)\nonumber\\
&&-3888 \bar{P}\left(5 \pi  \bar{Q}^2-2 \bar{S}\right)+729\big)=0.
\label{pt2}
\end{eqnarray}
The analytical solution of the above equations is not possible. However, we could find the critical points of the Kerr-Sen-AdS black holes by a dimensional analysis \cite{Cheng:2016bpx}
\begin{eqnarray}
\bar{P}_c = k_1(\epsilon) \cdot \bar{Q}^{-2},~~
\bar{S}_c =k_2(\epsilon)\cdot \bar{Q}^2,~~
\bar{T}_c = k_3(\epsilon) \cdot \bar{Q}^{-1},~~
\bar{G}_c =k_4(\epsilon) \cdot \bar{Q},~~
\bar{v}_c = k_5(\epsilon)\cdot \bar{Q},~~
\end{eqnarray}
The values of the various coefficients $k_1$, $k_2$, $k_3$, $k_4$ and $k_5$ can be found as follows
\begin{eqnarray}
k_1&=&\frac{-0.414674 \epsilon ^5-2.00863 \epsilon ^4-0.196446 \epsilon ^3+0.674347 \epsilon ^2+0.188842 \epsilon +0.00973587}{-145.155 \epsilon ^6-758.574 \epsilon ^5-336.022 \epsilon ^4+216.348 \epsilon ^3+155.856 \epsilon ^2+26.4646 \epsilon +1.04499},\nonumber\\
k_2&=&\frac{-18150. \epsilon ^6-298041. \epsilon ^5-417996. \epsilon ^4-81697.5 \epsilon ^3+30079.9 \epsilon ^2+6490.92 \epsilon +128.467}{-631.988 \epsilon ^5-10257.1 \epsilon ^4-12609.7 \epsilon ^3-669.172 \epsilon ^2+936.807 \epsilon +83.3434},\nonumber\\
k_3&=&\frac{143.6\epsilon^{3/2}+3.08 \epsilon ^{5/2}+0.004 \epsilon ^{7/2}-0.179\epsilon ^3-28.09\epsilon ^2-391.17 \epsilon +445.77 {\epsilon^{1/2}}-9.42}{-9401.8 \epsilon ^{3/2}-670.72 \epsilon ^{5/2}-4.29 \epsilon ^{7/2}+0.10 \epsilon ^4+73.57 \epsilon ^3+3433.0\epsilon ^2+11092.1 \epsilon -1715.28 {\epsilon^{1/2}}+1914.14},\nonumber\\
k_4&=&\frac{0.508\epsilon ^{3/2}+1.551\epsilon ^{5/2}+1.363\epsilon ^{7/2}+0.813\epsilon ^3+0.771\epsilon ^2+0.172\epsilon+0.046 {\epsilon}^{1/2}+0.006}{0.607\epsilon ^{3/2}+1.197 \epsilon^{5/2}+2.0066 \epsilon ^3+1.399\epsilon ^2+0.26266 \epsilon +0.0634 {\epsilon}^{1/2}+0.0086},\nonumber\\
k_5&=&\frac{405.143 \epsilon ^{3/2}+1705.67 \epsilon ^{5/2}+1946.96 \epsilon ^{7/2}+3484.3 \epsilon ^3+1821.21 \epsilon ^2+204.961 \epsilon +19.236 {\epsilon}^{1/2}+1.793}{245.824 \epsilon ^{3/2}+576.169 \epsilon ^{5/2}+321.952 \epsilon ^3+251.132 \epsilon ^2+47.742\epsilon +18.991 {\epsilon}^{1/2}+1.163}. 
\end{eqnarray}
We plot the coefficients $k_1$, $k_2$, $k_3$, $k_4$ and $k_5$ as functions of $\epsilon$ in \ref{Error}, where we find that the numerical values of these functions are highly consistent with their fitting curves.
As special cases, we examine the limits of the parameter $\epsilon\to 0$ and $ \infty$ to determine the limiting values. In the limit $\epsilon\to 0$, the critical values of various thermodynamic quantities are evaluated to be
\begin{eqnarray}
\bar{P}_c|_{\epsilon\to 0}&=&k_1(\epsilon)|_{\epsilon\to 0}\cdot \bar{Q}^{-2}=0.00931672\, \bar{Q}^{-2}, \nonumber\\
\bar{S}_c|_{\epsilon\to 0}&=&k_2(\epsilon)|_{\epsilon\to 0}\bar{Q}^{2}=1.54142\,\cdot \bar{Q}^{2},\nonumber\\
\bar{T}_c|_{\epsilon\to 0}&=&k_3(\epsilon)|_{\epsilon\to 0}\cdot \bar{Q}^{-1}=0.00492096\,\bar{Q}^{-1},\nonumber\\
\bar{G}_c|_{\epsilon\to 0}&=&k_4(\epsilon)|_{\epsilon\to 0}\cdot \bar{Q}=0.707256\,\bar{Q},\nonumber\\
\bar{v}_c|_{\epsilon\to 0}&=&k_5(\epsilon)|_{\epsilon\to 0}\,\cdot \bar{Q}=1.54256\,\bar{Q}. 
\end{eqnarray}
On the other hand, the critical points as $\epsilon\to\infty$ reduce to the values of the Kerr-AdS black holes \cite{Wei:2015ana}
\begin{eqnarray}
\bar{P}_c|_{\epsilon\to \infty}&=&k_1(\epsilon)|_{\epsilon\to \infty}\cdot \bar{Q}^{-2}=0.00285678\epsilon^{-1}\, \cdot \bar{Q}^{-2}\approx 0.0029 \bar{J}^{-1},\nonumber\\
\bar{S}_c|_{\epsilon\to \infty}&=&k_2(\epsilon)|_{\epsilon\to \infty}\bar{Q}^{2}=28.7189\epsilon\,\cdot \bar{Q}^{2}\approx 28.7189 \bar{J},\nonumber\\
\bar{T}_c|_{\epsilon\to \infty}&=&k_3(\epsilon)|_{\epsilon\to \infty}\cdot \bar{Q}^{-1}={0.0418263}\epsilon^{-1/2}\,\cdot \bar{Q}^{-1}\approx 0.0418 \bar{J}^{-1/2}, \nonumber\\
\bar{G}_c|_{\epsilon\to \infty}&=&k_4(\epsilon)|_{\epsilon\to \infty}\cdot \bar{Q}=0.679336\epsilon^{1/2}\,\cdot \bar{Q}\approx 0.6793 \bar{J}^{1/2}, \nonumber\\
\bar{v}_c|_{\epsilon\to \infty}&=&k_5(\epsilon)|_{\epsilon\to \infty}\,\cdot \bar{Q}=6.04736\epsilon^{1/2}\,\cdot \bar{Q}\approx{6.04736}\bar{J}^{1/2}. 
\end{eqnarray}
\subsection{Phase diagrams and coexistence curves}
\begin{figure}
    \includegraphics[height=0.3\textwidth]{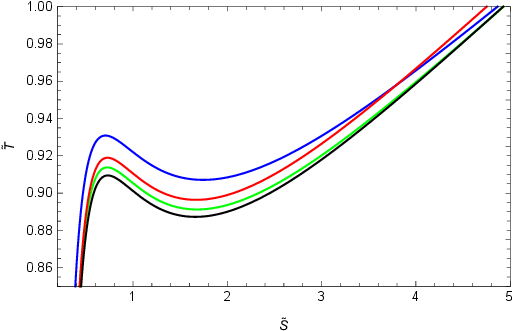}\hspace{0.2cm}
    \includegraphics[height=0.3\textwidth]{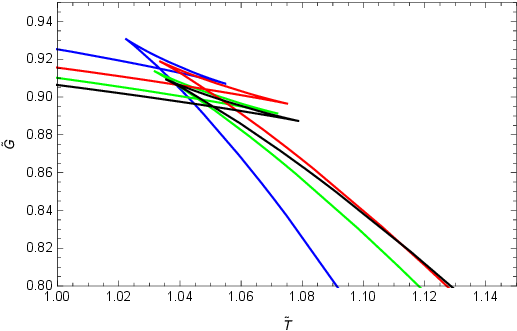}\\
    \includegraphics[height=0.3\textwidth]{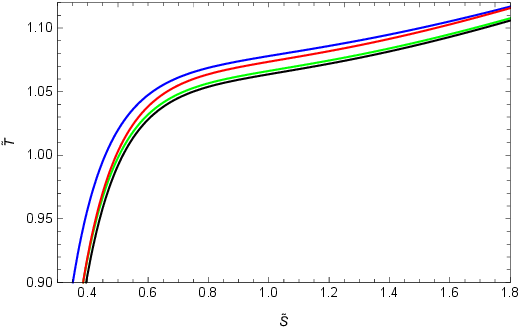}\hspace{0.2cm}
    \includegraphics[height=0.3\textwidth]{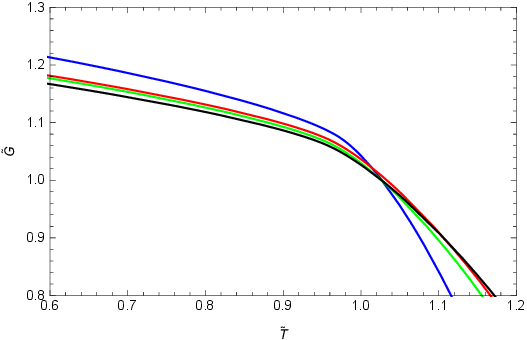}
    \caption{Plot showing the behaviors of $\widetilde{G}-\widetilde{T}$ and $\widetilde{T}-\widetilde{S}$ diagram depending on various values of the parameter $\epsilon$.  The upper panel we take the reduced pressure $\widetilde{P}=0.8(<\widetilde{P}_c=1.0)$ and in the lower panel $\widetilde{P}=1.2(>\widetilde{P}_c=1.0)$}
    \label{TS-GT} 
\end{figure}

The analysis of thermal phase transitions in asymptotically AdS black holes is crucial, particularly in light of their resemblance to a van der Waals-like fluid. Motivated by the various exotic properties and (un)stable phases of the AdS black holes, in our present work, we investigate the phase diagrams and coexistence curves of the Kerr-Sen-AdS black holes. We express the dimensionless temperature $\widetilde T$ and Gibbs free energy $\widetilde G$ in terms of the dimensionless pressure, entropy and $\epsilon$ as 
\begin{eqnarray}
\widetilde{T} &=&\frac{64 k_1^2 k_2^4 \widetilde{P}^2 \widetilde{S}^4+32 k_1 k_2^3 \widetilde{P} \widetilde{S}^3+k_2^2 \widetilde{S}^2 \left(16 \pi  k_1 \widetilde{P}+3\right)-12 \pi ^2 \epsilon ^2}{4 \sqrt{\pi } k_3 \left(k_2 \widetilde{S}\right){}^{3/2} \sqrt{\left(8 k_1 k_2 \widetilde{P} \widetilde{S}+3\right) \left(8 k_1 k_2^3 \widetilde{P} \widetilde{S}^3+3 k_2^2 \widetilde{S}^2+6 \pi  k_2 \widetilde{S}+12 \pi ^2 \epsilon ^2\right)}},\\
\widetilde{G} &=&\frac{12 \pi  k_2 \widetilde{S} \left(16 \pi  k_1 \epsilon ^2 \widetilde{P}+3\right)-64 k_1^2 k_2^4 \widetilde{P}^2 \widetilde{S}^4+3 k_2^2 \widetilde{S}^2 \left(16 \pi  k_1 \widetilde{P}+3\right)+108 \pi ^2 \epsilon ^2}{12 \sqrt{\pi } k_4 \sqrt{k_2 \widetilde{S}} \sqrt{\left(8 k_1 k_2 \widetilde{P} \widetilde{S}+3\right) \left(8 k_1 k_2^3 \widetilde{P} \widetilde{S}^3+3 k_2^2 \widetilde{S}^2+6 \pi  k_2 \widetilde{S}+12 \pi ^2 \epsilon ^2\right)}},\,\\
\widetilde{V}&=&\frac{4 \sqrt{k_1 \widetilde{S}} \left(8 k_1^4 \widetilde{P} \widetilde{S}^3+3 k_1^2 \widetilde{S}^2+3 \pi  k_1 \widetilde{S}+6 \pi ^2 \epsilon ^2\right)}{3 \sqrt{\pi } k_5 \sqrt{12 \pi ^2 \epsilon ^2 \left(8 k_1^2 \widetilde{P} \widetilde{S}+3\right)+\left(k_1 \widetilde{S} \left(8 k_1^2 \widetilde{P} \widetilde{S}+3\right)+3 \pi \right){}^2}}.
\end{eqnarray}
At this juncture, it is fascinating to show the expression of the horizon radius in terms of $k_1$ and $k_2$ as 
\begin{eqnarray}
{r}_{hc}=\frac{\sqrt{k_2} \left(\sqrt{\left(k_2 \left(8 k_1 k_2+3\right)+3 \pi \right){}^2}-3 \pi \right) Q}{\sqrt{\pi } \sqrt{\left(8 k_1 k_2+3\right) \left(8 k_1 k_2^3+3 k_2^2+6 \pi  k_2+12 \pi ^2 \epsilon ^2\right)}}.
\end{eqnarray}
Therefore, one can re-evaluate the horizon radius in a dimensionless way as
$$
\widetilde{r}=\frac{\sqrt{\left(8 k_1 k_2+3\right) \left(8 k_1 k_2^3+3 k_2^2+6 \pi  k_2+12 \pi ^2 \epsilon ^2\right)} \sqrt{k_2 \widetilde{S}} \left(\sqrt{\left(k_2 \widetilde{S} \left(8 k_1 k_2 \widetilde{P} \widetilde{S}+3\right)+3 \pi \right){}^2}-3 \pi \right)}{\sqrt{k_2} \left(\sqrt{\left(k_2 \left(8 k_1 k_2+3\right)+3 \pi \right){}^2}-3 \pi \right) \sqrt{\left(8 k_1 k_2 \widetilde{P} \widetilde{S}+3\right) \left(8 k_1 k_2^3 \widetilde{P} \widetilde{S}^3+3 k_2^2 \widetilde{S}^2+6 \pi  k_2 \widetilde{S}+12 \pi ^2 \epsilon ^2\right)}}.$$
Alternatively, the horizon radius in the reduced parameter space can be expressed as
\begin{eqnarray}
r_h=\frac{\sqrt{k_2 \widetilde{S}} \left(\sqrt{\left(k_2 \widetilde{S} \left(8 k_1 k_2 \widetilde{P} \widetilde{S}+3\right)+3 \pi \right){}^2}-3 \pi \right)}{\sqrt{\pi } \sqrt{\left(8 k_1 k_2 \widetilde{P} \widetilde{S}+3\right) \left(8 k_1 k_2^3 \widetilde{P} \widetilde{S}^3+3 k_2^2 \widetilde{S}^2+6 \pi  k_2 \widetilde{S}+12 \pi ^2 \epsilon ^2\right)}}.
\end{eqnarray}
We depict in \ref{TS-GT} the diagram for the reduced temperature $\widetilde{T}$ and the Gibbs free energy $\widetilde{G}$ with different $\epsilon$ values. The plots in the upper panel for $\widetilde{P}=0.8(<\widetilde{P}_c)=1.0$, show the behaviours of the reduced temperature pertaining to the oscillatory nature and the reduced Gibbs free energy expressing its swallow tail characteristics. It is observed that with the increase in the values of $\epsilon$, the local maxima and minima of the reduced temperature $\widetilde{T}$ decrease. The similar features for the intersecting points in the reduced Gibbs free energy $\widetilde{G}$ versus the reduced temperature plot are also reflected, where with the increase in the values of $\epsilon$ the shifting is observed in the lower $\widetilde{T}$ values and higher $\widetilde{G}$ values. Suddenly, for the reduced pressure, $\widetilde{P}=1.2(>\widetilde{P}_c)=1.0$, the oscillatory pattern in the $\widetilde{T}-\widetilde{S}$ diagram disappears, and similarly is the case for the $\widetilde{G}-\widetilde{T}$ diagram, where the swallow tail characteristics vanish as well. Hence, these diagrams are highly sensitive to the parameter $\epsilon$.

As the oscillatory nature of the isobaric lines in the $\widetilde{T}-\widetilde{S}$ planes, as well as the swallow tail characteristics of the $\widetilde{G}-\widetilde{T}$ diagram affirms to us that there should be a first-order small to large black holes phase transition. This occurs when the reduced pressure reaches values below its critical point. At the critical points, we have the coexistence line, where both small and large black hole phases coexist. This can be obtained either by establishing the Maxwell-equal area law on the $\widetilde{T}-\widetilde{S}$ plane or by evaluating the intersection point in the $\widetilde{G}-\widetilde{T}$ diagram.
\begin{figure}
\centering
    \includegraphics[height=0.4\textwidth]{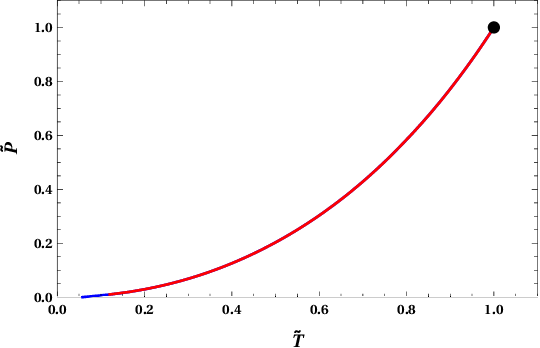}
     \caption{Coexistence curves in the $\widetilde{P}-\widetilde{T}$ plane for the Kerr-Sen-AdS black hole}
     \label{KNPT}
\end{figure}
All these characteristics on the $\widetilde{T}-\widetilde{S}$ or $\widetilde{G}-\widetilde{T}$ diagrams are solely determined with the values of $\epsilon$, rather than $\bar{J}$ and $\bar{Q}$. We solve numerically for each $\epsilon$, the coexistence curve for the small and large black hole phases. We apply the curve-fitting method to obtain the coexistence curve, where the small and large black hole phases coexist. Using the fitting formula \cite{Cheng:2016bpx}
\begin{eqnarray}
\widetilde P= \sum^{10}_{i=0}a_i \widetilde T^i,~~ \widetilde T \in (0,1)\,,
\end{eqnarray}
where the fitting coefficients $a_0$ to $a_{10}$ are function of the parameter $\epsilon=\bar{J}/\bar{Q}^2$. These results are utilised in analysing the thermodynamic properties of the Kerr-Sen-AdS black holes by varying them along the coexistence curve. By the fitting method, we construct the coexistence curve for $\epsilon=1.0$, which is expressed as 
\begin{eqnarray}
&~&\widetilde P=-122.519 \widetilde{T}^{10}+640.962 \widetilde{T}^9-1446.6 \widetilde{T}^8+1847.03 \widetilde{T}^7-1471.1 \widetilde{T}^6+760.614 \widetilde{T}^5\nonumber\\
&&-257.452 \widetilde{T}^4+56.3662 \widetilde{T}^3-6.83291 \widetilde{T}^2+0.559273 \widetilde{T}-0.0174441.
\end{eqnarray}
Next, we classify the phase transition behaviours as the first or higher orders by checking the Clausius-Clapeyron and the Ehrenfest equations. This analogy follows directly from the classical thermodynamic systems. One experiences a first-order phase transition when the Clausius-Clapeyron equations are satisfied, whereas for a second-order phase transition, we refer to the Ehrenfest equations. The identification of such characteristics is completely determined through the following equations \cite{Mo:2014mba, Ali:2019myr,Ali:2019mkl}
\begin{eqnarray}
\label{CCE}
\left(\frac{\partial \widetilde{P}}{\partial \widetilde{T}}\right)_{\widetilde{S}}=\frac{\widetilde{C}_{\widetilde{P_2}}-\widetilde{C}_{\widetilde{P_1}}}{\widetilde{V}\widetilde{T}\left(\widetilde{\alpha_2}-\widetilde{\alpha_1}\right)}=\frac{\Delta \widetilde{C}_{\widetilde{P}}}{\widetilde{V}\widetilde{T}\Delta\widetilde{\alpha}},
\end{eqnarray}
\begin{eqnarray}\label{EE}
\left(\frac{\partial \widetilde{P}}{\partial \widetilde{T}}\right)_{\widetilde{V}}=\frac{\widetilde{\alpha_2}-\widetilde{\alpha_1}}{\widetilde{\kappa}_{\widetilde{T_2}}-\widetilde{\kappa}_{\widetilde{T_1}}}=\frac{\Delta\widetilde{\alpha}}{\Delta\widetilde{\kappa}_{\widetilde{T}}},
\end{eqnarray}
where the quantities $\widetilde{\alpha}=\left(\frac{\partial \widetilde{V}}{\partial \widetilde{T}}\right)_{\widetilde{P}}/\widetilde{V}$ and $\widetilde{\kappa}_{\widetilde{T}}=-\left(\frac{\partial \widetilde{V}}{\partial \widetilde{P}}\right)_{\widetilde{T}}/\widetilde{V},$ correspond, respectively, to the volume expansion coefficient and the isothermal compressibility.
The expressions in \ref{tempads} and \ref{Volume} for temperature and volume are used to calculate the specific heat at constant pressure $\widetilde{C}_{\widetilde{P}}$, the volume expansion coefficient $\widetilde{\alpha}$ and the isothermal compressibility $\widetilde{\kappa}_{\widetilde{T}}$ of the rotating Kerr-Sen-AdS black holes
\begin{eqnarray}
\label{CP}
\widetilde{C}_P&=&\widetilde{T}\left(\frac{\partial \widetilde{S}}{\partial \widetilde{T}}\right)_{\widetilde{P}}=\bigg[2 k_2 \widetilde{S} \left(8 k_1 k_2 \widetilde{P} \widetilde{S}+3\right) \left(k_2 \widetilde{S} \left(k_2 \widetilde{S} \left(8 k_1 k_2 \widetilde{P} \widetilde{S}+3\right)+6 \pi \right)+12 \pi ^2 \epsilon ^2\right)\times\nonumber\\
&&\left(k_2^2 \widetilde{S}^2 \left(16 k_1 \widetilde{P} \left(2 k_2 \widetilde{S} \left(2 k_1 k_2 \widetilde{P} \widetilde{S}+1\right)+\pi \right)+3\right)-12 \pi ^2 \epsilon ^2\right)\bigg]{B(\epsilon,\widetilde{S},\widetilde{P})}^{-1}.
\end{eqnarray}
\begin{eqnarray}
\label{alpha}
\widetilde{\alpha}&=&\Bigg[12 \sqrt{\pi } \widetilde{Q} \left(k_1 \widetilde{S}\right){}^{3/2} \left(\left(8 k_1^2 \widetilde{P} \widetilde{S}+3\right) \left(k_1 \widetilde{S} \left(k_1 \widetilde{S} \left(8 k_1^2 \widetilde{P} \widetilde{S}+3\right)+6 \pi \right)+12 \pi ^2 \epsilon ^2\right)\right){}^{3/2}\nonumber\\
&&\big(6 \pi ^2 \epsilon ^2 \left(3 k_1^2 \widetilde{S}^2 \left(8 k_1^2 \widetilde{P} \widetilde{S}+3\right){}^2+2 \pi  k_1 \widetilde{S} \left(8 k_1^2 \widetilde{P} \widetilde{S}+9\right)+3 \pi ^2\right)+k_1 \widetilde{S} \left(k_1 \widetilde{S} \left(8 k_1^2 \widetilde{P} \widetilde{S}+3\right)+3 \pi \right){}^3\nonumber\\
&+&72 \pi ^4 \epsilon ^4\big)\Bigg]\Bigg[\left(k_1 \widetilde{S} \left(k_1 \widetilde{S} \left(8 k_1^2 \widetilde{P} \widetilde{S}+3\right)+3 \pi \right)+6 \pi ^2 \epsilon ^2\right) \big(12 \pi ^2 \epsilon ^2 \left(8 k_1^2 \widetilde{P} \widetilde{S}+3\right)+\big(k_1 \widetilde{S} \left(8 k_1^2 \widetilde{P} \widetilde{S}+3\right)\nonumber\\
&&+3 \pi\big){}^2\big)B(\epsilon,\widetilde{S},\widetilde{P})\Bigg]^{-1}.
\end{eqnarray}
\begin{eqnarray}
\label{kappa}
&&\widetilde{\kappa}_T=24 k_2 \widetilde{Q}^2 \widetilde{S} \Bigg[\big(-\big(6 \pi ^2 k_2 \epsilon ^2 \widetilde{S} \left(3 k_2 \widetilde{S} \left(8 k_1 k_2 \widetilde{P} \widetilde{S}+3\right){}^2+2 \pi  \left(8 k_1 k_2 \widetilde{P} \widetilde{S}+9\right)\right)+k_2^2 \widetilde{S}^2 \big(k_2^2 \widetilde{S}^2\nonumber\\
&&\left(8 k_1 k_2 \widetilde{P} \widetilde{S}+3\right){}^3+9 \pi  k_2 \widetilde{S} \left(8 k_1 k_2 \widetilde{P} \widetilde{S}+3\right){}^2+12 \pi ^2 \left(4 k_1 k_2 \widetilde{P} \widetilde{S}+3\right)\big)+72 \pi ^4 \epsilon ^4\big){}^2\nonumber\\
&&-3 \pi ^2 \left(k_2 \widetilde{S}+2 \pi  \epsilon ^2\right){}^2B(\epsilon,\widetilde{S},\widetilde{P})\big)\Bigg]\Bigg[B(\epsilon,\widetilde{S},\widetilde{P})\left(8 k_1 k_2 \widetilde{P} \widetilde{S}+3\right) \big(k_2 \widetilde{S} \left(k_2 \widetilde{S} \left(8 k_1 k_2 \widetilde{P} \widetilde{S}+3\right)+3 \pi \right)\nonumber\\
&&+6 \pi ^2 \epsilon ^2\big) \left(k_2 \widetilde{S} \left(k_2 \widetilde{S} \left(8 k_1 k_2 \widetilde{P} \widetilde{S}+3\right)+6 \pi \right)+12 \pi ^2 \epsilon ^2\right)\Bigg]^{-1}.
\end{eqnarray}
where the form of the function $B(\epsilon,\widetilde{S},\widetilde{P})$ is evaluated to be
\begin{eqnarray}
\label{Bform}
&&B(\epsilon,\widetilde{S},\widetilde{P})=144 \pi ^4 \epsilon ^4 \left(32 k_1^2 \widetilde{P} \widetilde{S}+9\right)+24 \pi ^2 k_1 \epsilon ^2 \widetilde{S} \big(k_1 \widetilde{S} \left(16 k_1^2 \widetilde{P} \widetilde{S}+3\right) \left(8 k_1^2 \widetilde{P} \widetilde{S}+3\right){}^2\nonumber\\
&&+36 \left(4 \pi  k_1^2 \widetilde{P} \widetilde{S}+\pi \right)\big)+k_1^4 \widetilde{S}^4 \big(32 k_1 \widetilde{P} \big(4 k_1 \widetilde{P} \big(k_1 \widetilde{S} \big(k_1 \widetilde{S} \left(16 k_1 \widetilde{P} \left(2 k_1 \widetilde{S} \left(k_1^2 \widetilde{P} \widetilde{S}+1\right)+3 \pi \right)+9\right)\nonumber\\
&&+36 \pi \big)-6 \pi ^2\big)+27 \pi \big)-27\big).
\end{eqnarray}
At the critical points the quantity $B(\epsilon,\widetilde{S},\widetilde{P})=B(\epsilon,k_1,k_2)$, has the value
\begin{eqnarray}
B(\epsilon,k_1,k_2)&=&1536 \pi  k_1^2 k_2^5 \left(16 \pi  k_1 \epsilon ^2+3\right)+3 k_2^4 \left(256 \pi ^2 k_1^2 \left(30 \epsilon ^2-1\right)+288 \pi  k_1-9\right)+6912 \pi ^2 k_1 k_2^3 \epsilon ^2\nonumber\\
&\;&+216 \pi ^2 \left(16 \pi  k_1+3\right) k_2^2 \epsilon ^2+288 \pi ^3 k_2 \epsilon ^2 \left(16 \pi  k_1 \epsilon ^2+3\right)+4096 k_1^4 k_2^8+4096 k_1^3 k_2^7\nonumber\\
&\;&+384 k_1^2 \left(16 \pi  k_1+3\right) k_2^6+1296 \pi ^4 \epsilon ^4. 
\end{eqnarray}
In the derivation of \ref{kappa}, we have utilized the thermodynamic identity
\begin{equation}
\label{identity}
\left(\frac{\partial \widetilde{V}}{\partial \widetilde{P}}\right)_{\widetilde{T}}=\left(\frac{\partial \widetilde{V}}{\partial \widetilde{P}}\right)_{\widetilde{S}}+\left(\frac{\partial \widetilde{V}}{\partial \widetilde{S}}\right)_{\widetilde{P}} \left(\frac{\partial \widetilde{S}}{\partial \widetilde{P}}\right)_{\widetilde{T}}.
\end{equation}%
It is noticeable that $\widetilde{C}_{\widetilde{P}}, \widetilde{\alpha}, \text{and}\, \widetilde{\kappa}_{\widetilde{T}}$ share
the same factor $B(\epsilon,\widetilde{S},\widetilde{P})$ in their denominators, and hints us the fact that they all diverge at same set of values of $\widetilde{S}, \text{and}\,\widetilde{P}$ corresponding to the critical points.\\

Let us now remark on the validity of Ehrenfest's equations,~\ref{CCE} and \ref{EE}, at the critical point. The definition of the volume expansion coefficient $\widetilde{\alpha}$ suggests that
\begin{equation}
\widetilde{V}\widetilde{\alpha}=\left(\frac{\partial \widetilde{V}}{\partial \widetilde{T}}\right)_{\widetilde{P}}=\left(\frac{\partial
\widetilde{V}}{\partial \widetilde{S}}\right)_{\widetilde{P}}\left(\frac{\partial \widetilde{S}}{\partial
\widetilde{T}}\right)_{\widetilde{P}}=\left(\frac{\partial \widetilde{V}}{\partial \widetilde{S}}\right)_{\widetilde{P}}\left(\frac{\widetilde{C}_{\widetilde{P}}
}{\widetilde{T}}.\right).
\label{CCE1}
\end{equation}
Therefore, the R.H.S. of \ref{CCE} can now be transformed into
\begin{equation}
\frac{\Delta \widetilde{C}_{\widetilde{P}}}{\widetilde{T}\widetilde{V}\Delta \widetilde{\alpha}}=\left[\left(\frac{\partial \widetilde{S}}{\partial
\widetilde{V}}\right)_{\widetilde{P}}\right]_c.
\label{CCE2}
\end{equation}
We use the expression for entropy in \ref{entropy}, the volume in \ref{volume_rh}, to get \ref{CCE2} as
\begin{equation}
\frac{\Delta \widetilde{C}_{\widetilde{P}}}{\widetilde{T}\widetilde{V}\Delta \widetilde{\alpha}}=\frac{k_3\sqrt{\pi k_2}\left(\left(8 k_1 k_2+3\right) \left(k_2 \left(k_2 \left(8 k_1 k_2+3\right)+6 \pi \right)+12 \pi ^2 \epsilon ^2\right)\right){}^{3/2}}{C(k_1,k_2,\epsilon)},
\label{323}
\end{equation}%
where
\begin{eqnarray}
C(k_1,k_2,\epsilon)&=&\big(9 k_2^4 \left(128 \pi ^2 k_1^2 \epsilon ^2+48 \pi  k_1+3\right)+3 \pi  k_2^3 \left(16 k_1 \left(18 \pi  \epsilon ^2+\pi \right)+27\right)\nonumber\\
&\;&+6 \pi ^2 k_2^2 \left(16 \pi  k_1 \epsilon ^2+27 \epsilon ^2+6\right)+108 \pi ^3 k_2 \epsilon ^2+512 k_1^3 k_2^7+576 k_1^2 k_2^6\nonumber\\
&\;&+72 k_1 \left(8 \pi  k_1+3\right) k_2^5+72 \pi ^4 \epsilon ^4\big).
\end{eqnarray}

Further, we use \ref{tempads}, to get the L.H.S. of \ref{CCE} as
\begin{equation}
\left[\left(\frac{\partial \widetilde{P}}{\partial \widetilde{T}}\right)_{\widetilde{S}}\right]_c=\frac{k_3\sqrt{\pi k_2}\left(\left(8 k_1 k_2+3\right) \left(k_2 \left(k_2 \left(8 k_1 k_2+3\right)+6 \pi \right)+12 \pi ^2 \epsilon ^2\right)\right){}^{3/2}}{C(k_1,k_2,\epsilon)}.
\label{325}
\end{equation}
From \ref{323} and \ref{325}, it is verified that the first of the Ehrenfest relations, \ref{CCE}, is equally valid. 

Keep on using \ref{alpha} and \ref{kappa} for the volume expansion coefficient and isothermal compressibility, respectively. The L.H.S of \ref{EE} is written as
\begin{eqnarray}
\left[\left(\frac{\partial \widetilde{P}}{\partial \widetilde{T}}\right)_{\widetilde{V}}\right]_c&=&\frac{\sqrt{\pi k_2} \left(\left(8 k_1 k_2+3\right) \left(8 k_1 k_2^3+3 k_2^2+6 \pi  k_2+12 \pi ^2 \epsilon ^2\right)\right){}^{3/2}}{D(k_1,k_2,\epsilon)}
\nonumber
\\
&\;& \times  \big(6 \pi ^2 \left(3 k_2^2 \left(8 k_1 k_2+3\right){}^2+2 \pi  k_2 \left(8 k_1 k_2+9\right)+3 \pi ^2\right) \epsilon ^2\nonumber\\
&\:&+k_2 \left(k_2 \left(8 k_1 k_2+3\right)+3 \pi \right){}^3+72 \pi ^4 \epsilon ^4\big)
,\label{326}
\end{eqnarray}%
where
\begin{eqnarray}
D(k_1,k_2,\epsilon)&=&2 \big(48 \pi ^4 k_2 \left(k_2 \left(64 k_1^2 k_2^2+64 k_1 k_2+15\right)+18 \pi \right) \epsilon ^4+24 \pi ^2 k_2^2 \big(k_2^2 \left(8 k_1 k_2+3\right){}^3\nonumber\\
&\;&+\pi  k_2 \left(192 k_1^2 k_2^2+176 k_1 k_2+39\right)+18 \pi ^2\big) \epsilon ^2+k_2^3 \big(k_2^3 \left(8 k_1 k_2+3\right){}^4+12 \pi  k_2^2 \left(8 k_1 k_2+3\right){}^3\nonumber\\
&\;&+96 \pi ^2 k_2 \left(16 k_1^2 k_2^2+14 k_1 k_2+3\right)+72 \pi ^3\big)+576 \pi ^6 \epsilon ^6\big).
\label{327}
\end{eqnarray}
 Again, we have used the thermodynamic identity, \ref{identity} while deriving \ref{326}. Now we use the definitions of isothermal compressibility coefficient $\widetilde{\kappa}_{\widetilde{T}}$ and volume expansion coefficient $\widetilde{\alpha}$, to derive the following relation
\begin{equation}
\widetilde{V}\widetilde{\kappa}_{\widetilde{T}}=-\left(\frac{\partial \widetilde{V}}{\partial \widetilde{P}}\right)_{\widetilde{T}}=\left(\frac{\partial
\widetilde{T}}{\partial \widetilde{P}}\right)_{\widetilde{V}}\left(\frac{\partial \widetilde{V}}{\partial
\widetilde{T}}\right)_{\widetilde{P}}=\left(\frac{\partial \widetilde{T}}{\partial \widetilde{P}}\right)_{\widetilde{V}}\widetilde{V}\widetilde{\alpha},\label{328}
\end{equation}%
using which we can derive the R.H.S of \ref{EE} as
\begin{eqnarray}
\frac{\Delta \widetilde{\alpha}}{\Delta \widetilde{\kappa}_{\widetilde{T}}}=\left[\left(\frac{\partial
\widetilde{P}}{\partial \widetilde{T}}\right)_{\widetilde{V}}\right]_c&=&\frac{\sqrt{\pi k_2} \left(\left(8 k_1 k_2+3\right) \left(8 k_1 k_2^3+3 k_2^2+6 \pi  k_2+12 \pi ^2 \epsilon ^2\right)\right){}^{3/2}}{D(k_1,k_2,\epsilon)}
\nonumber
\\
&\;& \times  \big(6 \pi ^2 \left(3 k_2^2 \left(8 k_1 k_2+3\right){}^2+2 \pi  k_2 \left(8 k_1 k_2+9\right)+3 \pi ^2\right) \epsilon ^2\nonumber\\
&\:&+k_2 \left(k_2 \left(8 k_1 k_2+3\right)+3 \pi \right){}^3+72 \pi ^4 \epsilon ^4\big).\label{329}
\end{eqnarray}%
Once again, we have used the thermodynamic identity \ref{identity} to derive \ref{329}. It is observed that \ref{329} reveals the validity of Ehrenfest equations. So far, we have proved that both the Ehrenfest equations are correct at the critical points.

Now using \ref{323} and \ref{329}, one can obtain
 \begin{eqnarray}
\frac{\Delta \widetilde{C}_{\widetilde{P}}}{\widetilde{T}\widetilde{V}\Delta \widetilde{\alpha}}-\frac{\Delta \widetilde{\alpha}}{\Delta \widetilde{\kappa}_{\widetilde{T}}}&=&\frac{1}{2} \frac{\sqrt{\pi k_2} \sqrt{\left(8 k_1 k_2+3\right) \left(8 k_1 k_2^3+3 k_2^2+6 \pi  k_2+12 \pi ^2 \epsilon ^2\right)}B(k_1,k_2,\epsilon)}{C(k_1,k_2,\epsilon)D(k_1,k_2,\epsilon)}
.\label{330}
\end{eqnarray}%
Since $B(k_1,k_2,\epsilon)$ vanishes at the critical point, one can infer that
 \begin{equation}
\frac{\Delta \widetilde{C}_{\widetilde{P}}}{\widetilde{T}\widetilde{V}\Delta \widetilde{\alpha}}-\frac{\Delta \widetilde{\alpha}}{\Delta \widetilde{\kappa}_{\widetilde{T}}}=0.\label{331}
\end{equation}%
Therefore, the \textit{ Prigogine-Defay} ratio is derived to be
\begin{equation}
\Pi=\frac{\Delta \widetilde{C}_{\widetilde{P}}\widetilde{\kappa}_{\widetilde{T}}}{\widetilde{T}\widetilde{V}\Delta \widetilde{\alpha}^2}=1.\label{332}
\end{equation}%
Hence, \ref{332} together with the validity of Ehrenfest equations enable us to conclude that the phase transition of the Kerr-Sen-AdS black holes in the extended phase space is a second-order one.

\section{Energy extraction from rotating Kerr-Sen-AdS black holes}
To calculate the energy extraction through the Penrose process, we need to investigate the internal energy, which is obtained by doing the Legendre transformation of the enthalpy, $H$, such that $U=H-PV$. In this section, we restore the dimensions of the quantities $S,P,J,Q$ to study the details of the Penrose process explicitly through efficiency. Now, to calculate the internal energy, we write the enthalpy in the form \cite{Dolan:2011xt} $$H=\sqrt{\alpha+\beta P+\gamma P^2},$$ so that the quantities $\alpha$, $\beta$, and $\gamma$ are expressed as
$$\alpha =\frac{36 \pi ^2 J^2+18 \pi  Q^2 S+9 S^2}{36 \pi  S},\,\,\beta =\frac{96 \pi ^2 J^2 S+48 \pi  Q^2 S^2+48 S^3}{36 \pi  S},\,\gamma=\frac{16 S^3}{9 \pi }.$$
Note that the discriminant $$\beta^2-4\alpha\gamma=\frac{16}{9} \left(2 \pi  J^2+Q^2 S\right)^2,$$ is a positive definite quantity. The volume is defined as \cite{Dolan:2011xt,Cvetic:2010jb} $$V=\left.\frac{\partial H}{\partial P}\right\vert_{S,J,Q}=\frac{1}{2}\frac{\beta+2\gamma P}{\sqrt{\alpha+\beta P+\gamma P^2}}=\frac{\beta+2\gamma P}{2H}\rightarrow P=\frac{2HV-\beta}{2\gamma}.$$ 
With above definitions, we can evaluate the enthalpy in terms of $V,S,J,Q$ as
\begin{eqnarray}
    \label{enthalp_new}
    H=\frac{1}{2}\sqrt{\frac{\beta^2-4\alpha\gamma}{V^2-\gamma}}=\frac{2 \sqrt{\pi } \sqrt{\left(2 \pi  J^2+Q^2 S\right)^2}}{\sqrt{9 \pi  V^2-16 S^3}}.
\end{eqnarray}
which immediately follows that $$V^2>\left(\frac{4\pi}{3}\right)^2\left(\frac{S}{\pi}\right)^3.$$ and in agreement with the result obtained in \cite{Dolan:2011xt}.\\
Clearly, the internal energy is simplified to
\begin{eqnarray}
    \label{internal_new}
U=H-PV=H-V\left(\frac{2HV-\beta}{2\gamma}\right)=-H\left(\frac{V^2}{\gamma}-1\right)+\frac{\beta V}{2\gamma}=\frac{\beta V}{2\gamma}-\frac{\sqrt{V^2-{\gamma}} \sqrt{{\beta}^2-4{\alpha} {\gamma}}}{2\gamma}.    
\end{eqnarray}
\begin{eqnarray}
    \label{internal_new01}
U=\frac{3 V \sqrt{9 \pi  V^2-16 S^3} \left(2 \pi ^2 J^2+S \left(\pi  Q^2+S\right)\right)+16 \sqrt{\pi } S^3 \left(2 \pi  J^2+Q^2 S\right)-9 \pi ^{3/2} V^2 \left(2 \pi  J^2+Q^2 S\right)}{8 S^3 \sqrt{9 \pi  V^2-16 S^3}}.
\end{eqnarray}
In the derivation of energy extraction, it would be more convenient to express the internal energy in terms of $S,P,J,Q$ as
\begin{eqnarray}
    \label{internal_new02}
    U=\frac{4 \pi ^2 J^2 (4 P S+3)+S \left(2 \pi  Q^2 (4 P S+3)+S (8 P S+3)\right)}{2 \sqrt{\pi } \sqrt{S} \sqrt{8 P S+3} \sqrt{12 \pi ^2 J^2+S \left(S (8 P S+3)+6 \pi  Q^2\right)}}.
\end{eqnarray}
Next, we discuss the efficiency of the Kerr-Sen-AdS black holes and calculate their values for different limiting cases. The efficiency is calculated by taking the ratio of the mechanical energy extracted in a physical process and the initial heat energy (enthalpy)
$$\eta=\frac{W_{\text{max}}}{M_i}.$$
where $M_i$ is the initial mass of the black hole, and for the AdS case, it is the enthalpy $H$. On the other hand, the mechanical energy could be identified as the difference between the initial and final internal energies. For isobaric and isentropic processes, the AdS black hole produces mechanical work and can be obtained by decreasing both the angular momentum and the charge. If $J$ and $Q$ are reduced from some non-zero values to zero, the efficiency is calculated to be
\begin{eqnarray}
    \label{internal_new03}
\eta=\frac{U(J,Q)-U(0,0)}{H(J,Q)}.
\end{eqnarray}
Beforehand, if we look into the expression of the thermodynamic volume, we can observe that it is a monotonically increasing function of $J$. Decreasing the value of $J$ will result in a monotonic increase in the volume, and hence the area remains constant. During the whole process, the black hole also heats up. One of the subtleties of a charged black hole is that it can extract work during isotropic and isentropic processes. On the one hand, one can decrease the value of $J$ keeping the charge $Q$ constant.
On the other hand, we keep the angular momentum $J$ fixed while increasing the value of $Q$. As a final possibility, we can decrease both $J$ and $Q$ to improve the efficiency of this process. The only change that matters is the initial and final values of $U$. \\
\begin{eqnarray}
    \label{internal_new04}
&&\eta=\frac{\pi  L^2 \left(2 \pi ^2 J^2 \left(2 \pi  L^2+S\right)+S \left(\pi  L^2 \left(2 \pi  Q^2+S\right)+S \left(\pi  Q^2+S\right)\right)\right)}{\left(\pi  L^2+S\right) \left(4 \pi ^3 J^2 L^2+\pi  L^2 S \left(2 \pi  Q^2+S\right)+S^3\right)}\nonumber\\
&&-\frac{\pi  L^2 S}{\sqrt{\left(\pi  L^2+S\right) \left(4 \pi ^3 J^2 L^2+\pi  L^2 S \left(2 \pi  Q^2+S\right)+S^3\right)}}.    
\end{eqnarray}
The maximal efficiency is obtained in the extremal limit where  temperature vanishes, and in such a case, we have the maximum value of the angular momentum, such that
$$J_\text{max}^2=\frac{S^2 \left(\pi ^2 L^4+2 \pi  L^2 \left(\pi  Q^2+2 S\right)+3 S^2\right)}{4 \pi ^4 L^4}.$$
which is a positive definite quantity. We can speculate that the efficiency in this limit would read as
\begin{eqnarray}
    \label{eta_00}
\eta=\frac{4 \pi ^2 L^4 \left(\pi  Q^2+S\right)+\pi  L^2 S \left(2 \pi  Q^2+7 S\right)+3 S^3}{2 \left(\pi  L^2+S\right) \left(2 \pi  L^2 \left(\pi  Q^2+S\right)+3 S^2\right)}-\frac{\pi ^{3/2} L^3 S}{\left(\pi  L^2+S\right) \sqrt{2 \pi  L^2 S \left(\pi  Q^2+S\right)+3 S^3}}.   
\end{eqnarray}
Several limiting cases for the efficiency are in order
\begin{eqnarray}
\left.\eta\right\vert_{S\to\infty}=\frac{1}{2}=50\%,\nonumber\\
\left.\eta\right\vert_{S\to0}=1=100\%.
\end{eqnarray}
In addition, in the limit $L\to\infty$, the efficiency of the rotating Kerr-Sen black hole turns out to be
\begin{eqnarray}
 \eta=\frac{-\sqrt{2} \sqrt{S \left(\pi  Q^2+S\right)}+2 \pi  Q^2+2 S}{2 \pi  Q^2+2 S}.
\end{eqnarray}
which in the non-charged limit ($Q=0$) reduces to
$$\left.\eta\right\vert_{Q\to0,L\to\infty}=29.3.\%$$
The detailed derivation of the Kerr-AdS or Kerr-Newman-AdS cases could be found in Ref. \cite{Dolan:2011xt}. The efficiency for different limiting values of the rotation parameter ($J$), the charge parameter ($Q$), entropy ($S$) and the curvature radius ($L$) is discussed in detail in this work. The efficiency of approximately $29\%$ obtained in the asymptotically flat case ($L\to \infty$) is the minimal one and is a well-known result \cite{Dolan:2011xt, Wald:1984rg}

\section{Adiabatic compressibility and velocity of sound}
The specific heat at constant pressure has been derived using the expression for the enthalpy. In this section, we derive the specific heat at constant volume using the expression for the internal energy. It reads as
\begin{eqnarray}
    C_V=T\left.\frac{\partial S}{\partial T}\right\vert_{V,J,Q}=T\left.\frac{1}{\frac{\partial^2U}{\partial S^2}}\right\vert_{V,J,Q}.
\end{eqnarray}
We can show explicitly that for any positive set of values of the parameters $P,S,Q,J$, ratio $\frac{C_P}{C_V}\geq 1$. The quantity $C_P$ diverges along the curve in the $JP$ versus $SP$ diagram, where $JP$ and $SP$ are the dimensionless quantities. 
\begin{figure}
    \centering
    \includegraphics[width=0.45\linewidth]{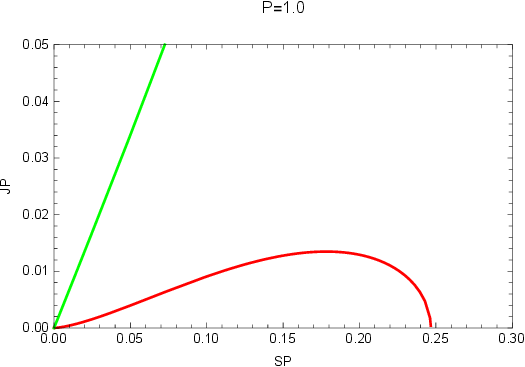}
      \includegraphics[width=0.45\linewidth]{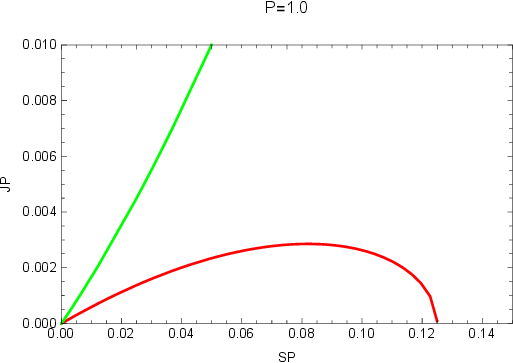}\\
    \caption{The graphs show the behaviour of the specific heat at constant pressure (red curve) and the curve when $T=0$ (i.e., $J=J_\text{max}$) in the $JP-SP$ plane. The region below the red curve shows that there $C_P$ is negative, while along the curve it diverges. On the other hand, the green shows that the Hawking temperature vanishes there, and consequently, the heat capacity also vanishes. The left figure shows the behaviour for the Kerr-Sen-AdS black holes, while the right figure depicts the behaviour for the Kerr-AdS back hole.}
    \label{fig:placeholder}
\end{figure}
The region below the red curve corresponds to $C_P<0$, and the critical values lie on the curve and correspond to the maximum value of $JP$ at $SP=0.0820434$ and hence $JP=0.00285678$ (for Kerr-AdS case). Likewise, for the Kerr-Sen-AdS black holes, we have the critical points at $JP=0.013482$ for $SP=0.178231$. When $Q=0$, the heat capacity reduces to
$$C_P=\frac{2 (8 P S+3) \left(64 P^2 S^2+16 P \left(\pi  Q^2+2 S\right)+3\right) \left(S (8 P S+3)+6 \pi  Q^2\right)}{4096 P^4 S^4+2048 P^3 S^2 \left(3 \pi  Q^2+2 S\right)-384 P^2 \left(2 \pi ^2 Q^4-12 \pi  Q^2 S-3 S^2\right)+864 \pi  P Q^2-27}.$$ which in addition for $J=0$ $$C_P=\frac{2 S (8 P S+1)}{8 P S-1}.$$ In this case the ratio ${C_P}/{C_V}$ reads as
$$\frac{C_P}{C_V}=\frac{1}{1-8PS}\geq1$$ and $PS$ lies in the range $0\leq PS<1/8$.\\
The adiabatic compressibility and the speed of sound have a complementary relationship among themselves. These quantities are fundamentally important when discussed in the context of astrophysics. The very definition of the adiabatic compressibility must be followed as:
\begin{eqnarray}
    \label{AC_basic}
    \kappa_S=-\frac{1}{V}\left(\frac{\partial V}{\partial P}\right)_{S,J}.
\end{eqnarray}
Using the expressions for volume in \ref{Volume}, we can calculate the adiabatic compressibility
\begin{eqnarray}
    \label{AC_1}
   \kappa_S=\frac{36 S \left(j^2+2 \pi  q^2\right)^2}{(8 p+3) \left(\left(3 j^2+8 p+3\right)+6 \pi  q^2\right) \left(\left(3 j^2+16 p+6\right)+6 \pi  q^2\right)}.    
\end{eqnarray}
This is manifestly positive for any arbitrary values of the rotation ($J$) and charge ($Q$) parameters. The quantities $q=Q/\sqrt{S}$, $j=2\pi J/S$ and $p=PS$ are dimensionless. In the limit when $Q=0$, we have the compressibility corresponding to that of the Kerr-AdS black holes. For both $J=0$ and $Q=0$, the compressibility vanishes, thereby the non-rotating black holes in the absence of charge are incompressible. When the angular momentum is increased, the compressibility also increases and attains its maximum value when $J=J_{\it max}$. There adiabatic compressibility also attains maximum at $J_{\text{ max}}$
\begin{eqnarray}
    \label{AC_max}
\kappa_S\bigg|_{T=0}=\frac{2 S \left(8 p+2 \pi  q^2+1\right)^2}{(8 p+3) \left(4 p+\pi  q^2+1\right) \left((8 p+3) +2 \pi  q^2\right)}.  
\end{eqnarray}
This value also corresponds to the extremal case when the temperature vanishes. The speed of sound, $v_s$, in the adiabatic process is associated in the thermodynamic sense such that
\begin{eqnarray}
    \label{speed AD}
    v_s^{-2}=\frac{\partial \rho}{\partial P}\bigg|_{S,J}=1+\rho\kappa_S=1+\frac{9 \left(j^2+2 \pi  q^2\right)^2}{\left(\left(3 j^2+16 p+6\right)+6 \pi  q^2\right)^2}.
\end{eqnarray}
where $\rho=\frac{M}{V}$ is the density. Although we define the speed of sound in the thermodynamic sense, it is not associated with the kind of surface that generates across the event horizon as restricted by the famous no-hair theorem \cite{Dolan:2011jm}. Nevertheless, it may be associated with a specific breathing mode for Kerr-Sen-AdS black holes, as the pressure changes with volume while the area remains constant. The speed of sound in~\ref{speed AD} has the values in the range $0\leq v_s\leq 1$ for any values of the rotation and charge parameters. Although for the Kerr black hole the speed of sound approaches unity, it is not the case for Kerr-Sen-AdS black holes ($Q\neq 0$). In the extremal limit we have, $T=0$, and hence $J=J_{\text{max}}$. In this limit, we have the speed of sound as
\begin{eqnarray}
\label{extremal Speed}
  v_s^{-2}=1+\frac{\left(8 p+2 \pi  q^2+1\right)^2}{\left(8 p+2 \pi  q^2+3\right)^2}.
\end{eqnarray}
Even in the limit $P=0$, $v_s$ has $q$-dependence, hence differing from the results of the Kerr-AdS case. In addition, if $Q=0$, then the value of $v_s^{-2}=10/9$, and we recover the results for Kerr-AdS case \cite{Dolan:2011jm}. \\ 
Keeping the maximum angular momentum in the extremal limit as constant, we get the quantity
\begin{eqnarray}
    \label{dSdP}
 \frac{dS}{dP}\bigg|_{T=0}=-\frac{8 S^2 \left(8 p+\pi  q^2+2\right)}{128 p^2+16 p \left(\pi  q^2+3\right)+3}.
\end{eqnarray}
Likewise, we have 
\begin{align*}
    \kappa_{T=0}=\frac{2 S \left(4 \pi ^2 (32 p+11) q^4+16 \pi  p (40 p+27) q^2+8 p (128 p (p+1)+41)+8 \pi ^3 q^6+68 \pi  q^2+33\right)}{\left(4 p+\pi  q^2+1\right) \left(8 p+2 \pi  q^2+3\right) \left(16 p \left(8 p+\pi  q^2+3\right)+3\right)}. 
\end{align*}
It should be mentioned that the speed of sound, as in \ref{speed AD} and \ref{extremal Speed}, is written in terms of the speed of light $c$. We recover the results for Kerr-AdS black holes when the charge parameter $Q=q=0$ \cite{Dolan:2011jm}. All the expressions for non-rotating AdS are also discussed in this work.

\section{A precursor to the holographic thermodynamics}
Despite its significant achievements, the thermodynamics of black holes in AdS spacetime, as well as the proper interpretation of their holographic dual from the perspective of black hole chemistry, has remained an enigma \cite{Mann:2024sru}. As far as the AdS/CFT correspondence is concerned, the cosmological constant $\Lambda$ ($<0$) sets the asymptotic geometry of AdS spacetime and is expressed as a relationship to the number of colours $N$ in its dual field theory \cite{ Maldacena:1997re, Karch:2015rpa}
\begin{equation}
    k\frac{L^{d-2}}{16\pi G}=N^2,
\end{equation}
where $L$ is the curvature radius of the AdS geometry and the factor $k$ is numerically determined from the details of the particular holographic system. Hence, the variation in the negative $\Lambda$ corresponds to the corresponding variation of the number of colours $N$. Such variations are accountable for changing the dual gauge field theories in the boundary \cite{Johnson:2014yja, Cong:2021jgb, Cong:2021fnf}.\\
The idea of the variable Newton's constant \cite{Cong:2021fnf} has been extended to a more general boundary metric with a boundary radius that qualitatively differs from the AdS radius. In this respect, Visser proposed that for the holographic description of the AdS black hole thermodynamics, one may require a chemical potential conjugate to its colour charge and hence extend the idea of the thermodynamic Euler equation to the high-energy states in the large-$N$ gauge theories \cite{Visser:2021eqk}. Several research activities have been devoted based on the idea \cite{Zheng:2024glr, Zhang:2023uay, Sadeghi:2023tuj, Ali:2023ppg, Punia:2023ilo}. However, from Visser's perspective, since the boundary radius does not coincide with the AdS radius, we can vary the conformal factor of the boundary CFT metric while keeping Newton's constant fixed. Subsequently, we may develop a boundary volume $\mathcal{V}$ and a central charge $C$ which are independent of each other. Later, in a follow-up paper by Ahmed et. al., the exact dictionary between the extended phase space and its holographic dual in the CFT was constructed \cite{Ahmed:2023snm}. Based on this proposal, a sufficient number of works have been done to investigate the holographic thermodynamics of AdS spacetimes, including the RN-AdS \cite{Ladghami:2024wkv}, Kerr-AdS \cite{Gong:2023ywu, Ahmed:2023dnh}, Kerr-Newman-AdS black holes \cite{Baruah:2024yzw}, and five-dimensional neutral Gauss-Bonnet AdS black hole \cite{Yang:2024krx}.  \\
\subsection{Incorporating Newton's constant in thermodynamic quantities}
In this section, we incorporate Newton's constant in the mass formula for the Kerr-Sen-AdS black holes as given in \ref{KSAdS}. In doing so, we restore Newton's gravitational constant $G_N$ in the definition of the entropy and the pressure \cite{Caldarelli:1999xj}. In the expressions of the mass formula, we denote the $\widetilde{M}$ by $M$ only, and we drop the overbar from various thermodynamic quantities. Therefore, the square of the mass term after restoring Newton's constant is written in the form
\begin{eqnarray}
    \label{massSq}
    M^2=\frac{8}{3} \pi  J^2 P G_N+\frac{\pi  J^2}{S G_N}+\frac{16 P^2 S^3 G_N^3}{9 \pi }+\frac{4}{3} P Q^2 S G_N^2+\frac{4 P S^2 G_N}{3 \pi }+\frac{S}{4 \pi  G_N}+\frac{Q^2}{2}.
\end{eqnarray}
which can be further reduced in terms of the area $A$ and the cosmological constant $\Lambda$ as
\begin{eqnarray}
    \label{massSqA}
    M^2=\frac{A^3 \Lambda ^2}{2304 \pi ^3 G_N^2}-\frac{A^2 \Lambda }{96 \pi ^2 G_N^2}+\frac{A}{16 \pi  G_N^2}+\frac{4 \pi  J^2}{A}-\frac{A \Lambda  Q^2}{24 \pi }-\frac{J^2 \Lambda }{3}+\frac{Q^2}{2}.
\end{eqnarray}
The first law and corresponding Smarr relation, while treating $G_N$ and $\Lambda$ as dynamical variables, are derived as
\begin{eqnarray}
    \label{firstlaw_1}
    &&dM=\frac{\kappa}{8\pi G_N}dA+\Phi dQ+\Omega dJ-V\frac{d\Lambda}{8\pi G_N}+(M-\Phi Q-\Omega J)\frac{dG_N}{G_N},\\
&&M=\frac{\kappa A}{4\pi G_N}+\Phi Q+2\Omega dJ+\frac{V\Lambda}{4\pi G_N}.
\end{eqnarray}
The related thermodynamic quantities could be derived as 
\begin{eqnarray}
    \label{thermoQuan}
 \frac{\kappa}{8\pi G_N}&=&\left(\frac{\partial{M}}{\partial{A}}\right)_{J,Q,\Lambda,G_N}=\frac{1}{2M}\left(\frac{A^2 \Lambda ^2}{768 \pi ^3 G_N^2}-\frac{4 \pi  J^2}{A^2}-\frac{A \Lambda }{48 \pi ^2 G_N^2}+\frac{1}{16 \pi  G_N^2}-\frac{\Lambda  Q^2}{24 \pi }\right),\nonumber\\
\Phi&=&\left(\frac{\partial{M}}{\partial{Q}}\right)_{J,A,\Lambda,G_N}=\frac{1}{2M}\left(Q-\frac{A \Lambda  Q}{12 \pi }\right),\nonumber\\
\Omega&=&\left(\frac{\partial{M}}{\partial{J}}\right)_{Q,A,\Lambda,G_N}=\frac{1}{2M}\left(\frac{8 \pi  J}{A}-\frac{2 J \Lambda }{3}\right),\nonumber\\
\frac{V}{8\pi G_N}&=&\left(\frac{\partial{M}}{\partial{\Lambda}}\right)_{J,Q,A,G_N}=\frac{1}{2M}\left(\frac{A^3 \Lambda }{1152 \pi ^3 G_N^2}-\frac{A^2}{96 \pi ^2 G_N^2}-\frac{A Q^2}{24 \pi }-\frac{J^2}{3}\right),\nonumber\\
\frac{(M-\Phi Q-\Omega J)}{G_N}&=&\left(\frac{\partial{M}}{\partial{G_N}}\right)_{J,Q,\Lambda,G_N}=\frac{1}{2M}\left(\frac{A^3 \Lambda }{1152 \pi ^3 G_N^2}-\frac{A^2}{96 \pi ^2 G_N^2}-\frac{A Q^2}{24 \pi }-\frac{J^2}{3}\right).
\end{eqnarray}

\subsection{Mixed thermodynamic quantities}
In this section, we incorporate the holographic dual relation of the AdS length $L$, Newton's constant $G_N$, and the central charge $C$ to express $C$ in a generic $ d$-dimensional as 
$$C=\frac{\Omega_{d-2}L^{d-2}}{16\pi G_N},$$
In our case $d=2$ and hence $C=\frac{L^2}{4G_N}$. Since $L^2=\frac{3}{8\pi G_N P}$, we could express it to write $G_N$ in terms of $C$ and $P$ as \cite{Qu:2022nrt, Sadeghi:2024ish} $$G_N=\frac{1}{4}\sqrt{\frac{3}{2\pi CP}}.$$
Now we write the mass term in terms of ($C,P,G_N,S, J, Q$) as follows
\begin{eqnarray}
    \label{mixedMass}
    M^2=\frac{S^3 \sqrt{C P}}{8 \sqrt{6} \pi ^{5/2} C^2}+\frac{4 \pi ^{3/2} \sqrt{\frac{2}{3}} J^2 \sqrt{C P}}{S}+\frac{\sqrt{\frac{2 \pi }{3}} J^2 \sqrt{C P}}{C}+\frac{S^2 \sqrt{C P}}{\sqrt{6} \pi ^{3/2} C}+\sqrt{\frac{2}{3 \pi }} S \sqrt{C P}+\frac{Q^2 S}{8 \pi  C}+\frac{Q^2}{2}.\nonumber\\
\end{eqnarray}

The corresponding first law may be written as
\begin{eqnarray}
    \label{mixedfirstlaw}
    dM=TdS+\Phi dQ+\Omega dJ+V dP+\mu dC.
\end{eqnarray}
The related thermodynamic quantities are expressed as
\begin{eqnarray}
T&=&\left(\frac{\partial{M}}{\partial{S}}\right)_{J,Q,P,C}=\frac{16 \pi ^2 \sqrt{6} C^2 \sqrt{C P} \left(S^2-4 \pi ^2 J^2\right)+2 \pi  C S^2 \left(8 \sqrt{6} S \sqrt{C P}+3 \sqrt{\pi } Q^2\right)+3 \sqrt{6} S^4 \sqrt{C P}}{96 \pi ^{5/2} C^2 M S^2},\\
\Phi&=&\left(\frac{\partial{M}}{\partial{S}}\right)_{J,S,P,C}=\frac{Q (4 \pi  C+S)}{8 \pi  C M}.\\
\Omega&=&\left(\frac{\partial{M}}{\partial{S}}\right)_{Q,S,P,C}=\frac{\sqrt{\frac{2 \pi }{3}} J P^{1/2} (4 \pi  C+S)}{M S \sqrt{C}},\\
V&=&\left(\frac{\partial{M}}{\partial{S}}\right)_{J,Q,S,C}=\frac{P \left(16 \pi ^2 C^2 \left(4 \pi ^2 J^2+S^2\right)+8 \pi  C S \left(2 \pi ^2 J^2+S^2\right)+S^4\right)}{32 \sqrt{6} \pi ^{5/2} M S (C P)^{3/2}},\\
\mu &=&\left(\frac{\partial{M}}{\partial{S}}\right)_{J,Q,S,P,C}=\frac{16 \pi ^2 \sqrt{6} C^2 P \left(4 \pi ^2 J^2+S^2\right)-8 \pi  \sqrt{6} C P S \left(2 \pi ^2 J^2+S^2\right)-3 \left(4 \pi ^{3/2} Q^2 S^2 \sqrt{C P}+\sqrt{6} P S^4\right)}{192 \pi ^{5/2} C^2 M S \sqrt{C P}}.\nonumber\\
\end{eqnarray}
The free energy is expressed as
\begin{eqnarray}
F&=&M-TS=\frac{16 C^2 \left(12 \pi ^4 \sqrt{2} J^2 \sqrt{C P}+\pi ^2 S \left(\sqrt{2} S \sqrt{C P}+\sqrt{3 \pi } Q^2\right)\right)}{8 \pi ^{5/4} C \sqrt{S (4 \pi  C+S) \left(2 C \left(8 \pi ^3 \sqrt{6} J^2 \sqrt{C P}+\pi  S \left(2 \sqrt{6} S \sqrt{C P}+3 \sqrt{\pi } Q^2\right)\right)+\sqrt{6} S^3 \sqrt{C P}\right)}}\nonumber\\
&+&\frac{2 C \left(16 \pi ^3 \sqrt{2} J^2 S \sqrt{C P}+\pi ^{3/2} \sqrt{3} Q^2 S^2\right)-\sqrt{2} S^4 \sqrt{C P}}{8 \pi ^{5/4} C \sqrt{S (4 \pi  C+S) \left(2 C \left(8 \pi ^3 \sqrt{6} J^2 \sqrt{C P}+\pi  S \left(2 \sqrt{6} S \sqrt{C P}+3 \sqrt{\pi } Q^2\right)\right)+\sqrt{6} S^3 \sqrt{C P}\right)}}.
\end{eqnarray}

%%%%%%%%%%%%%%%%%%%%%%%%%%%%%%%%%%%%%%%%%%%%%%%%%%%%%%%%%%%%%%%%%%%%%%%%%%%%%%%%%%%%%%%%%%%%%%%%%%%%%%%%%%%%%%%%%%%%%%%%%%%%%%%%%%%%%%%%%%%%%%%%%%%%%%%%%%%%%%%%%%%%%%%%%%%%%%%%%%
\section{CFT thermodynamics}
To find the CFT thermodynamics, we set the CFT metric to be of the form \cite{Gubser:1998bc,Witten:1998qj,Ahmed:2023snm,Savonije:2001nd,Baruah:2024yzw}
\begin{eqnarray}
    \label{cft_metric}
    ds^2=\omega^2(-dt^2+L^2 d\Omega_2^2),
\end{eqnarray}
The spatial volume of the boundary CFT may be written as
\begin{eqnarray}
    \mathcal{V}=\Omega_2 (\omega L)^2. 
\end{eqnarray}
Correspondingly, in the language of holography, we can write the boundary thermodynamic quantities. Hence, we have a pressure term $p$ corresponding to the volume $\mathcal{V}$, i.e., -$p d\mathcal{V}$. The first law of the CFT thermodynamics and its corresponding Euler relation may be cast as
\begin{eqnarray}
    \label{first law_Euler}
    dE&=&\mathcal{T}d\mathcal{S}+\mathcal{\varphi}d\mathcal{Q}-p d\mathcal{V}+\mu d{C}+\Omega d{\mathcal{J}}\nonumber\\
    E&=&\mathcal{T}\mathcal{S}+\mathcal{\varphi}\mathcal{Q}+\mu C+\Omega \mathcal{J}\nonumber\\
    p&=&-\frac{E}{2}\frac{1}{\mathcal{V}},
\end{eqnarray}
All the related thermodynamic quantities are evaluated to be of the form
\begin{eqnarray}
    \label{CFTTemp_01}
    E^2&=&\frac{(4 \pi  C+\mathcal{S}) \left(4 \pi  C \left(4 \pi ^2 \mathcal{J}^2+\mathcal{S}^2\right)+2 \pi ^2 \mathcal{Q}^2 \mathcal{S}+\mathcal{S}^3\right)}{16 \pi ^2 C \mathcal{S} \mathcal{V}},\\
    \label{CFTTemp_02}
\mathcal{T}&=&\frac{16 \pi ^2 C^2 \left(\mathcal{S}^2-4 \pi ^2 \mathcal{J}^2\right)+16 \pi  C \mathcal{S}^3+2 \pi ^2 \mathcal{Q}^2 \mathcal{S}^2+3 \mathcal{S}^4}{8 \pi  \sqrt{C \mathcal{S}^3 \mathcal{V} (4 \pi  C+\mathcal{S}) \left(4 \pi  C \left(4 \pi ^2 \mathcal{J}^2+\mathcal{S}^2\right)+2 \pi ^2 \mathcal{Q}^2 \mathcal{S}+\mathcal{S}^3\right)}}r,\\  
\label{CFTTemp_03}
\mathcal{\varphi}&=&\frac{\pi  \mathcal{Q} \sqrt{\mathcal{S}} (4 \pi  C+\mathcal{S})}{2 \sqrt{C \mathcal{V} (4 \pi  C+\mathcal{S}) \left(4 \pi  C \left(4 \pi ^2 \mathcal{J}^2+\mathcal{S}^2\right)+2 \pi ^2 \mathcal{Q}^2 \mathcal{S}+\mathcal{S}^3\right)}},\\
\label{CFTTemp_04}
p&=&\frac{C \mathcal{S} \sqrt{(4 \pi  C+\mathcal{S}) \left(4 \pi  C \left(4 \pi ^2 \mathcal{J}^2+\mathcal{S}^2\right)+2 \pi ^2 \mathcal{Q}^2 \mathcal{S}+\mathcal{S}^3\right)}}{8 \pi  (C \mathcal{S} \mathcal{V})^{3/2}},\\
\label{CFTTemp_05}
\mu&=&\frac{16 \pi ^2 C^2 \left(4 \pi ^2 \mathcal{J}^2+\mathcal{S}^2\right)-\mathcal{S}^2 \left(2 \pi ^2 \mathcal{Q}^2+\mathcal{S}^2\right)}{8 \pi  \sqrt{C^3 \mathcal{S} \mathcal{V} (4 \pi  C+\mathcal{S}) \left(4 \pi  C \left(4 \pi ^2 \mathcal{J}^2+\mathcal{S}^2\right)+2 \pi ^2 \mathcal{Q}^2 \mathcal{S}+\mathcal{S}^3\right)}},\\
\label{CFTTemp_06}
\Omega&=&\frac{4 \pi ^2 \sqrt{C} \mathcal{J} (4 \pi  C+\mathcal{S})}{\sqrt{\mathcal{S} \mathcal{V} (4 \pi  C+\mathcal{S}) \left(4 \pi  C \left(4 \pi ^2 \mathcal{J}^2+\mathcal{S}^2\right)+2 \pi ^2 \mathcal{Q}^2 \mathcal{S}+\mathcal{S}^3\right)}}.
\end{eqnarray}
In the above relations, \ref{CFTTemp_02} represents the equation of state of the CFT, whereas \ref{CFTTemp_01} encodes the information regarding the internal energy formula. Also, one can relate the bulk thermodynamics to the corresponding first law of the conformal field theory (CFT), \ref{first law_Euler}. We also evaluate the chemical potential conjugate to the central charge $C$. Also mention that while the cosmological constant $ \Lambda $ is taken into consideration, the variable Newton's constant is not there in the first law and the corresponding Smarr relation.

\section{Conclusions}\label{Conclusions}
In this paper, we study the horizon structure of the Kerr-Sen-AdS black holes and their thermal properties. We presented the expressions for the temperature, mass, and volume of black holes in terms of the horizon radius, AdS radius, and dilation-axion parameters. Further considering the cosmological constant as the thermal pressure and its conjugate quantity as the thermodynamic volume, we express the mass, temperature, volume, and Gibbs free energy in terms of the pressure, entropy, angular momentum, and charge parameter. Furthermore, to investigate the critical points, we calculated the critical values of the various thermodynamic quantities in terms of the dimensionless quantity $\epsilon$. All the critical quantities are reduced to the corresponding expressions for the Kerr-AdS black holes in the limit when $\epsilon\to\infty$. The analysis of the $\widetilde{P}-\widetilde{T}$ through the critical points showed us that there should exist two phases of the Kerr-Sen-AdS black holes. Further, we plotted the $\widetilde{T}-\widetilde{S}$ and $\widetilde{G}-\widetilde{T}$ diagrams to show the point of inflexion and the swallowtail behavior. To phases of the black holes coexist at the critical points, and analysis of which is carried out by applying the Maxwell equal-area law $\widetilde{T}-\widetilde{S}$ where oscillatory behaviours are observed or by evaluating the intersection points on the $\widetilde{G}-\widetilde{T}$ diagrams. We numerically found the coexistence line for the $\widetilde{P}-\widetilde{T}$ curve and curve-fitted it. \\
Furthermore, we analytically verified the Ehrenfest equations to determine  the phase transition at the critical point of $\widetilde{T}-\widetilde{S}$ criticality for rotating Kerr-Sen-AdS black holes. Additionally, we have computed the volume expansion coefficient $\widetilde{\alpha}$, the isothermal compressibility $\widetilde{\kappa}_{\widetilde{T}}$, and the specific heat at constant pressure $\widetilde{C}_{\widetilde{P}}$. We found that these numbers diverge at the crucial locations and have a common factor in their denominators. Using the quantities $\widetilde{C}_{\widetilde{P}}$, $\widetilde{\kappa}_{\widetilde{T}}$, $\widetilde{\alpha}$, and the thermodynamic identity in \ref{identity}, we verify both the  Ehrenfest equations analytically. To conclude that the rotating
Kerr-Sen-AdS black holes in four dimensions exhibited a second-order phase transition. We calculated the Prigogine-Defay ratio at the critical points and showed that it is exactly equal to unity. The results confirmed the universality class of the thermodynamic properties of the rotating Kerr-Sen-AdS as the same as those of van der Waals fluids. \\
In the framework of gauge/gravity duality or the AdS/CFT correspondence, studying critical phenomena in the extended phase space of black hole thermodynamics offers several intriguing implications. This is due to the fact that the dual CFT at finite temperature is described by thermodynamics in AdS spacetimes, namely the AdS/CFT correspondence. To know the nature of quantum gravity, the thermodynamics of black holes in asymptotically AdS spacetime is important, particularly in the context of the AdS/CFT correspondence. At the onset of extended phase space thermodynamics, more generally, the black hole chemistry, as well as the holographic dictionary of the first laws of black hole mechanics, does not directly translate to the thermodynamics of the holographic dual. Since variation of the cosmological constant ($\Lambda<0$) in the extended phase space corresponds to varying the central charge and the CFT volume. If the central charge is treated as a constant variable, the field theory also remains the same. In such a case, varying $\Lambda$ is subjected to changing the CFT volume as well as Newton's constant. Several studies have been conducted in this direction, where the mixing of variations in the pressure term and Newton's gravitational constant occurs \cite{Ali:2023ppg}. It would be very interesting to explore whether any exciting correlations exist between extended phase space thermodynamics and its holographic duality for the Kerr-Sen-AdS black hole as well.

\section*{Acknowledgments}
MSA and SGG would like to thank the Inter-University Centre for Astronomy and  Astrophysics (IUCAA), Pune for hospitality while part of this work was being done.

%%%%%%%%%%%%%%%%%%%%%%%%%%%%%%%%%%%%%%%%%%%%%%%%%%%%%%%%%%%%%%%%%%%%%%%%%%%%%%%%%%%%%%%%%

\end{document}